\documentclass{aastex7}
\usepackage[utf8]{inputenc}
\usepackage{graphicx} 
\usepackage{amsmath}
\usepackage{amssymb}
\usepackage{hyperref}
\usepackage{natbib}
\usepackage{xcolor}
\usepackage{comment}
\DeclareMathAlphabet{\altmathcal}{OMS}{cmsy}{m}{n}

\begin{document}

\title{Magnetohydrodynamic Precipitation}

\author[0000-0002-3514-0383]{G. Mark Voit}
\affiliation{Michigan State University,
Department of Physics and Astronomy,
East Lansing, MI 48824, USA}
\email{voit@msu.edu}

\author{Benjamin D. Wibking}
\affiliation{Michigan State University,
Department of Physics and Astronomy,
East Lansing, MI 48824, USA}
\email{wibkingb@msu.edu}

\author{Doruk Yaldiz}
\affiliation{Michigan State University,
Department of Physics and Astronomy,
East Lansing, MI 48824, USA}
\email{yaldizdo@msu.edu}

\begin{abstract}
Circumgalactic gas around massive galaxies generally has a volume-filling component---an atmosphere---with a temperature determined by the potential-well depth of the galaxy's halo. If the atmosphere is near hydrostatic equilibrium and is stable to convection, then it can remain nearly homogeneous, as long as it is not too dense. But if its density is great enough, it becomes prone to producing a rain of cold clouds that fall toward the galaxy's center and accrete onto its central black hole. Here we explain how relatively weak magnetic fields enhance a galactic atmosphere's tendency to produce cold clouds and how the cold gas becomes organized into vertically elongated, highly magnetized filaments descending at sub-Keplerian speeds. It is intended to complement recent numerical simulations of the process and to serve as a guide to interpreting both simulations and observations of the filamentary gas in hot galactic atmospheres.
\end{abstract}

\section{Introduction}

This tutorial article seeks to clarify how weak magnetic fields enable small density perturbations to grow to non-linear amplitudes in gravitationally stratified galactic atmospheres, resulting in what the article will call \textit{magnetohydrodynamic precipitation}. The word ``precipitation" in this context refers to gas clouds that are condensing out of a hotter and more diffuse atmosphere. As with raindrops, gravity pulls newly formed clouds of denser gas toward the center of the potential well confining the atmosphere, and momentum transfer from the clouds to the ambient gas slows their descent. Here we will be focusing on how momentum transfer mediated by magnetic fields assists precipitation of gas clouds in hot galactic atmospheres.

We are providing a tutorial because the existing literature on the subject is not easy to follow. The earliest efforts to assess how magnetic fields affect precipitation in galactic atmospheres were formal thermal stability analyses considering only the initial stages of cloud growth, while the perturbation amplitudes remain small \cite[e.g.,][]{Loewenstein_1990,Balbus_1991}. Those classic results can be expressed in terms of a seventh-order polynomial dispersion relation, because seven equations describe the system, but the high-order polynomial does not lucidly reveal the system's seven normal modes. Later numerical simulations followed cloud growth into the nonlinear regime \citep[e.g.,][]{Ji_2018,Wibking_2025MNRAS.544.2577W}. But as we will see, interpretations of those simulations require insight into how \textit{nonlinear} effects either inhibit or enhance growth of the system's normal modes.

To prioritize clarity, we will therefore introduce complexity into the tutorial one step at a time, so that readers can digest the physical significance of each new addition. Here is how the tutorial proceeds:
\begin{itemize}

    \item Section \ref{sec:ThermalInstability} introduces the mathematical formalism used to describe thermal instability and derives an equation for modeling thermal instability in stratified galactic atmospheres.

    \item Section \ref{sec:Buoyancy} discusses how buoyancy couples with thermal instability in a stratified atmosphere. It shows that small-amplitude entropy perturbations are usually thermally unstable and can excite internal gravity waves ($g$-modes) in the galactic atmospheres we would like to understand.
    
    \item Section \ref{sec:MagAccel} introduces the formalism we will use to account for how magnetic forces act upon thermally unstable blobs of atmospheric plasma.
    
    \item Section \ref{sec:MTD1} shows that magnetic forces add a new thermally unstable mode, distinct from the thermally unstable $g$-modes, in atmospheres with initially uniform horizontal magnetic fields. It calls the new mode \textit{magnetothermal drip} because the descending gas blobs that form rely on magnetic fields for support as they cool.
    
    \item Section \ref{sec:Saturation} considers how magnetothermal drip progresses as the density contrast of a thermally unstable plasma blob becomes nonlinear: The blob tends to elongate into a vertically oriented filament that descends at a speed significantly smaller than the potential well's circular velocity $v_\mathrm{c}$ and amplifies the local magnetic field strength as it descends.
    
    \item Section \ref{sec:PressurePerts} temporarily sets aside magnetic forces and considers pressure perturbations, which sections \ref{sec:Buoyancy} and \ref{sec:MTD1} ignore. It focuses on pressure perturbations because they are crucial for understanding how magnetothermal drip develops in atmospheres with magnetic fields that are not horizontal.
    
    \item Section \ref{sec:MTD2} adds pressure perturbations to the analysis of section \ref{sec:MTD1} and shows how magnetothermal drip develops in atmospheres with \textit{any} magnetic-field orientation.
    
    \item Section \ref{sec:MHDPrecip} builds upon section \ref{sec:MTD2} to describe how magnetohydrodynamic precipitation proceeds from magnetothermal drip and illustrates how precipitation develops in numerical simulations.
    
    \item Section \ref{sec:Torques} draws attention to an intriguing feature of magnetothermal drip in spherical potential wells: The magnetic fields it amplifies apply torques that aid accretion of condensed gas onto the central black hole, helping to fuel an active galactic nucleus (AGN) that heats the atmosphere.

    \item Section \ref{sec:Conduction} evaluates how thermal conduction would affect thermal stability, if included in the analysis.
    
    \item Section \ref{sec:Summary} summarizes the tutorial's main points.

\end{itemize}
An appendix discusses topics that would have disrupted the tutorial's flow if stuffed into the main narrative: Section \ref{app:gMode_Saturation} of the appendix explains why nonlinear saturation can suppress growth of thermally unstable $g$-modes before precipitation can develop.  Section \ref{app:Freefall} points out that thermal instability in a background medium near freefall does not saturate. Section \ref{app:AcousticStability} considers the thermal stability of acoustic modes and shows that they either remain stable or grow slowly. Section \ref{app:EntropyMode} links magnetothermal drip with the ``entropy mode" of magnetohydrodynamics. Section \ref{app:PreviousWork} discusses how the tutorial's analysis relates to previous work on thermal instability in magnetized galactic atmospheres.

\section{Generic Thermal Instability}
\label{sec:ThermalInstability}

We begin by reviewing the basic features of thermal instability in a diffuse, optically thin astrophysical medium. Later sections will build on this foundation. First, we assume that the pressure $P$ and density $\rho$ of the thermally unstable medium satisfy the following equation of state:
\begin{equation}
    P = K \rho^\gamma\; \; .
\end{equation}
Here, $\gamma$ is the usual adiabatic index and the logarithm of $K$ is proportional to the medium's specific entropy $s$. Isobaric density perturbations then correspond to entropy perturbations. We will consider the medium thermally unstable if the amplitudes of low-entropy (high-density) perturbations grow because those regions cool more quickly, or gain heat more slowly, than unperturbed fluid elements.

Thermal stability of low-entropy perturbations consequently depends on how the medium's net cooling rate changes in response to its thermodynamic state. Suppose the medium gains thermal energy per unit mass at the rate $\mathcal{H}$ and loses it at the rate $\mathcal{C}$. Its net heating rate per particle is then
\begin{equation}
    \frac {dq} {dt}
        \: = \: \mu m_p 
                \left( \altmathcal{H} - \altmathcal {C} \right) 
        = - \mu m_p \altmathcal{L}
\end{equation}
in which $\altmathcal{L} \equiv \altmathcal {C} - \altmathcal{H}$ represents net cooling and $\mu m_p$ is the mean mass per particle. Net heating is related to the change in specific entropy through $dq / ds = kT$, and so specific entropy changes at the rate
\begin{equation}
    \frac {ds} {dt} 
        \: = \: \frac {1} {kT} \frac {dq} {dt}
        \: = \: - \frac {\mu m_p  \altmathcal{L}} {kT}
    \; \; .
\end{equation}
We therefore find
\begin{equation}
    \frac {d \ln K} {dt} 
        \: = \: ( \gamma - 1) \frac {ds} {dt} 
        \: = \: - ( \gamma - 1 )  \frac {\altmathcal{L} \rho} {P}
        \: = \: - \frac {\altmathcal{L} \rho} {u}
    \; \; ,
\end{equation}
in which the right-hand side applies to a medium with thermal pressure $P = \rho k T / \mu m_p$ and thermal energy density $u = P / (\gamma -1)$. 

To make the formalism more intuitive, we can define both a cooling time $t_{\rm cool}$ and a heating time $t_{\rm heat}$:
\begin{equation}
    t_{\rm cool} \equiv \frac {u} {\altmathcal{C} \rho} 
    \; \; , \; \; 
    t_{\rm heat} \equiv \frac {u} {\altmathcal{H} \rho} 
    \; \; .
\end{equation}
With those definitions, a fluid element's entropy change over time can be expressed as
\begin{equation}
    \frac {d \ln K} {dt} 
        = \frac {1} {t_{\rm heat}} - \frac {1} {t_{\rm cool}} 
    \; \; .
\end{equation}
Entropy contrasts among neighboring fluid elements can therefore grow if either $t_{\rm heat}$ or $t_{\rm cool}$ is spatially inhomogeneous. And if both timescales are similar, then entropy contrasts grow at a rate similar to $t^{-1}_{\rm cool}$, regardless of whether the atmosphere as a whole is gaining or losing heat energy.

Now consider how an entropy perturbation in such a medium evolves. Suppose a fluid element initially located at $\mathbf{r}_0$ starts with a value of $K$ slightly different from the value $K_0$ it would have if there were no perturbation. We can represent that entropy difference using the symbol
\begin{equation}
    \Delta_K (\mathbf{r}_0,t) 
        \equiv \ln \left[ \frac {K(\mathbf{r}_0,t)} {K_0(\mathbf{r}_0,t)} \right]
    \; \; .
\end{equation}
The entropy difference $\Delta_K$ between a perturbed fluid element and its counterpart in an unperturbed atmosphere then evolves according to 
\begin{equation}
    \dot{\Delta}_K 
        \: \equiv \: \frac {d} {dt} \Delta_K
        \: = \: \frac {d \ln K} {dt} - \frac {d \ln K_0} {dt} 
    \; \; .
\end{equation}
However, when modeling thermal instability, we are more interested in the evolution of the \textit{local} entropy contrast at a particular location $\mathbf{r}$: 
\begin{equation}
    \delta_K (\mathbf{r},t) \equiv \ln \left[ \frac {K(\mathbf{r},t)} {K_0(\mathbf{r},t)} \right]
    \; \; .
\end{equation}
We therefore need to account for how the initial location $\mathbf{r}_0$ of a perturbed fluid element with coordinates $(\mathbf{r},t)$ may differ from the initial location of an unperturbed fluid element with those same coordinates.

\begin{figure}[t]
    \centering
    \includegraphics[width=0.99\linewidth]{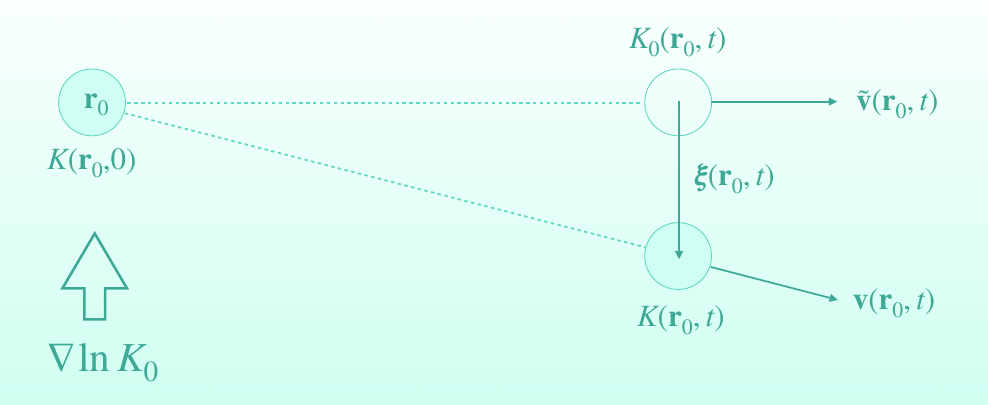}
    \caption{Schematic illustration of an evolving entropy perturbation in a stratified atmosphere. Shading represents the background entropy gradient $\nabla \ln K_0$, which is vertical. A perturbation initially at location $\mathbf{r}_0$ begins with an entropy difference $\Delta_K$ relative to the background atmosphere at that location. Unperturbed gas starting at the same place would have moved with velocity $\tilde{\mathbf{v}}$ but the perturbed gas moves with velocity $\mathbf{v}$. The perturbation's displacement $\boldsymbol{\xi}$ relative to the unperturbed flow therefore changes with time, and so its entropy contrast $\delta_K$ relative to neighboring gas depends on evolution of both the \textit{intrinsic} entropy difference $\Delta_K$ and the \textit{extrinsic} entropy change $\boldsymbol{\xi} \cdot \nabla \ln K_0$.}
    \label{fig:Schematic1}
\end{figure}

To account for differences in the initial locations of fluid elements, let $\mathbf{v}(\mathbf{r}_0,t)$ represent the trajectory of a perturbed fluid element that starts at $\mathbf{r}_0$, and let $\mathbf{\tilde{v}}(\mathbf{r}_0,t)$ represent the trajectory of an \textit{unperturbed} fluid element starting at the same place (see Figure \ref{fig:Schematic1}). We then find
\begin{equation}
    \frac {\partial} {\partial t} \delta_K  
        = \dot{\Delta}_K
           - ( \mathbf{v} - \mathbf{\tilde{v}} ) 
                \cdot \nabla \ln K_0
           - \mathbf{v} \cdot \nabla \delta_K
           \; \; .
           \label{eq:partial_deltaK}
\end{equation}
The first term on the right is the only one that depends on local cooling or heating. The second one depends on the displacement $\boldsymbol{\xi}(\mathbf{r}_0,t)$ between the trajectory of a perturbed fluid element and its unperturbed counterpart, and the displacement evolves according to 
\begin{equation}
    \boldsymbol{\dot{\xi}}(\mathbf{r}_0,t) = \mathbf{v}(\mathbf{r}_0,t) - \mathbf{\tilde{v}}(\mathbf{r}_0,t)
    \; \; .
\end{equation}
Displacements of perturbed fluid elements relative to the background entropy gradient $\nabla \ln K_0$ change the local entropy contrast $\delta_K$ without changing $\Delta_K$, and that is what the second term in equation (\ref{eq:partial_deltaK}) captures. The third term accounts for the flow of entropy perturbations through the location where the time derivative of $\delta_K$ is being evaluated. We can minimize the third term's significance by analyzing thermal instability in a frame comoving with the unperturbed flow. In that frame, the rate of change of $\delta_K$ is
\begin{equation}
    \dot{\delta}_K 
        = \dot{\Delta}_K 
            - \boldsymbol{\dot{\xi}} \cdot \nabla \ln K_0
            - \boldsymbol{\dot{\xi}} \cdot \nabla \delta_K
            \; \; .
\end{equation}
The third term then scales as the square of the perturbation amplitude and can be ignored in a linearized analysis. Hereafter, the symbol $\dot{\delta}_K$ will implicitly represent changes in local entropy contrast in the frame of an unperturbed flow. 

Distinguishing between $\dot{\Delta}_K$ and $\dot{\delta}_K$ becomes important when we want to assess how thermal stability depends on local conditions. In an optically thin galactic atmosphere, the local cooling time is generally a function of density $\rho$ and atmospheric temperature $T$, meaning that it can also be expressed as a function of $P$ and $K$. Taking into account just the local pressure and entropy perturbations, $\delta_P = \ln (P/P_0)$ and $\delta_K = \ln (K/K_0)$, we can express changes in the entropy difference $\Delta_K$ using the equation
\begin{equation}
    \dot{\Delta}_K = \omega_{\rm ti} \delta_K + \omega_{\rm sw} \delta_P
\end{equation}
along with the definitions
\begin{eqnarray}
    \omega_{\rm ti}
        & \equiv & \frac {1} {t_{\rm cool}}
                    \left( \frac {\partial \ln t_{\rm cool}} {\partial \ln K} 
                           - \frac {t_{\rm cool}} {t_{\rm heat}}
                                \frac {\partial \ln t_{\rm heat}} {\partial \ln K}    \right)_P
        \\
    \omega_{\rm sw}
        & \equiv & \frac {1} {t_{\rm cool}}
                    \left( \frac {\partial \ln t_{\rm cool}} {\partial \ln P} 
                           - \frac {t_{\rm cool}} {t_{\rm heat}}
                                \frac {\partial \ln t_{\rm heat}} {\partial \ln P}    \right)_K
                                \; \; .
\end{eqnarray}
The quantity $\omega_{\rm ti}$ is the rate at which the amplitudes of isobaric entropy perturbations evolve, and they are thermally unstable for $\omega_{\rm ti} > 0$. The quantity $\omega_{\rm sw}$ is the analogous rate for isentropic pressure perturbations (i.e.~sound waves), but it is typically negative, meaning that isentropic pressure perturbations are usually stable and decay with time. Both rates are similar in magnitude to $1/t_{\rm cool}$, as long as heating does not overwhelm cooling.

The equation we will use to model thermal instability in galactic atmospheres is therefore
\begin{equation}
    \dot{\delta}_K 
        = \omega_{\rm ti} \delta_K - \boldsymbol{\dot{\xi}} \cdot \nabla \ln K_0
        \; \; .
        \label{eq:TI-nabla}
\end{equation}
It represents the rate at which local entropy contrasts of linear order evolve in a frame comoving with an unperturbed atmospheric flow. The first term on the right-hand side describes changes in entropy contrast coming from local gains or losses of heat energy. The second one describes changes in entropy contrast that arise from a perturbed fluid element's motion relative to an entropy gradient in the unperturbed flow. 

\section{Buoyancy}
\label{sec:Buoyancy}

Close attention to vertical displacements becomes necessary when thermal instability couples with buoyancy. If a galactic atmosphere is convectively stable, then its entropy gradient $\nabla \ln K_0$ points in a direction opposite to the effective gravitational acceleration. Differentiating equation (\ref{eq:TI-nabla}) with respect to time then gives
\begin{equation}
    \ddot{\delta}_K - \omega_{\rm ti} \dot{\delta}_K = - \frac {\ddot{\xi}_\parallel} {\lambda_K}
        \label{eq:TI}
        \; \; ,
\end{equation}
where $\xi_\parallel$ is the component of $\boldsymbol{\xi}$ parallel to the entropy gradient and $\lambda_K \equiv | \nabla \ln K_0 |^{-1}$ is the local entropy scale height. Technically, the dots over the $\delta_K$ symbols represent partial derivatives with respect to time at a location comoving with the unperturbed flow, while the dots over $\xi_\parallel$ represent total time derivatives. However, the correction term needed to account for the difference between $\ddot{\xi}_\parallel$ and the partial time derivative of $\dot{\xi}_\parallel$ is second-order, and so we are ignoring it. We are also choosing to ignore time-dependent changes in $\omega_\mathrm{ti}$ and $\lambda_K$.

To model thermal instability in a magnetized atmosphere, we will eventually need to make use of the momentum equation for magnetohydrodynamics:
\begin{equation}
    \frac {d \mathbf{v}} {dt}
        = \mathbf{g}  - \frac {\nabla P} {\rho} 
            + \frac {(\nabla \times \mathbf{B}) \times \mathbf{B}}
                    {4 \pi \rho}
        \; \; .
\end{equation}
Here, $\mathbf{g}$ represents gravitational acceleration and $\mathbf{B}$ represents the magnetic field. To keep the notation compact, we will represent the magnetic component of acceleration with 
\begin{equation}
    \mathbf{a_B} \equiv \frac {(\nabla \times \mathbf{B}) \times \mathbf{B}} {4 \pi \rho}
                \label{eq:MagneticAcceleration}
\end{equation}
when we are not explicitly working with magnetic fields.

This tutorial will consider cases in which magnetic acceleration of the unperturbed flow is negligible. The momentum equation for the unperturbed flow is then
\begin{equation}
    \frac {d \mathbf{\tilde{v}}} {dt} 
        = \mathbf{g} - \frac {\nabla P_0} {\rho_0}
        \; \; .
\end{equation}
Subtracting it from the momentum equation for the perturbed flow, making use of the relation $\delta_\rho \equiv \ln(\rho/\rho_0) = (\delta_P - \delta_K) / \gamma$, and dropping terms that are second-order in the perturbation amplitudes then gives us a  momentum equation for displacements:
\begin{equation}
    \boldsymbol{\ddot{\xi}} 
        = - \frac {\mathbf{\tilde{g}}} {\gamma}
                    \left[ \delta_K + (\gamma - 1) \delta_P \right]
            - \frac {c_{\rm s}^2} {\gamma} \nabla \delta_P
            + \mathbf{a_B}
            \label{eq:xi_momentum}
            \; \; .
\end{equation}
Here, $c_{\rm s}^2 \equiv \gamma P_0 / \rho_0$ is the sound speed in the unperturbed flow, and 
\begin{equation}
    \mathbf{\tilde{g}} 
        \equiv \mathbf{g} - \frac {d \mathbf{\tilde{v}}} {dt}
        = \frac {\nabla P_0} {\rho_0}
    \label{eq:gtilde}
\end{equation}
is the effective gravitational acceleration in a frame comoving with the unperturbed flow.

To illustrate how entropy perturbations in a stratified atmosphere drive displacements, and vice versa, we will temporarily simplify equation (\ref{eq:xi_momentum}) by setting both $\delta_P$ and $\mathbf{a_B}$ to zero. The equation governing thermal instability then reduces to
\begin{equation}
    \ddot{\delta}_K - \omega_{\rm ti} \dot{\delta}_K + \tilde{\omega}_{\rm BV}^2 {\delta_K} = 0
        \; \; ,
\end{equation}
where $\tilde{\omega}_{\rm BV}^2 \equiv |\mathbf{\tilde{g}}| / \gamma \lambda_K$ is the Brunt V\"{a}is\"al\"a frequency for buoyant oscillations of adiabatic $g$-modes in a frame with an effective gravitational acceleration $\mathbf{\tilde{g}}$ \citep{Vaisala_1925SSFCP...2...19V,Brunt_1927QJRMS..53...30B}. Perturbations that have $\delta_K \propto e^{-i \omega t}$ must satisfy
\begin{equation}
        \omega^2 - i \omega_{\rm ti} \omega - \tilde{\omega}_{\rm BV}^2 = 0
        \; \; .
\end{equation}
The frequency solutions are then
\begin{equation}
    \omega_\pm \equiv \pm \, \tilde{\omega}_{\rm BV} 
            \left( 1 - \frac {\omega_{\rm ti}^2} {4 \tilde{\omega}_{\rm BV}^2} \right)^{1/2}
            + \frac {i \omega_{\rm ti}} {2} 
            \; \; .
            \label{eq:omega_pm}
\end{equation}
According to equation (\ref{eq:omega_pm}), isobaric entropy perturbations are thermally unstable for $\omega_{\rm ti} > 0$, and their amplitudes oscillate (i.e. they are $g$-modes) if $\tilde{\omega}_{\rm BV} > \omega_{\rm ti}/2$.

\begin{figure}[t]
    \centering
    \includegraphics[width=0.99\linewidth]{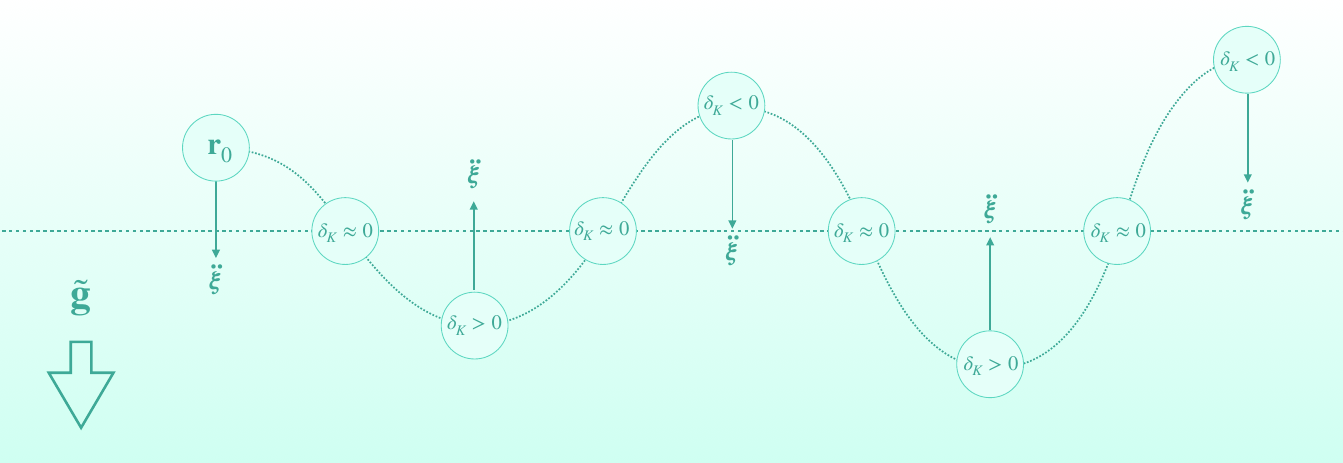}
    \caption{Schematic illustration of a thermally unstable entropy perturbation in a stratified atmosphere with $\tilde{\omega}_\mathrm{BV} > \omega_\mathrm{ti}$. Shading represents the background entropy gradient, which increases in a direction opposite to the effective gravitational acceleration $\mathbf{\tilde{g}}$. A low-entropy perturbation initially at location $\mathbf{r}_0$ accelerates downward and overshoots the layer at which its entropy contrast is zero $(\delta_K = 0)$. Its entropy contrast then becomes positive, and so it starts to accelerate upward, resulting in oscillations with a frequency approximately equal to $\tilde{\omega}_\mathrm{BV}$. As $\delta_K$ oscillates, the perturbation's intrinsic entropy difference $\Delta_K$ also oscillates, with an amplitude ratio $|\Delta_K|/|\delta_K| \sim \omega_\mathrm{ti}/\tilde{\omega}_\mathrm{BV}$, and thermal pumping causes the amplitudes of both $\delta_K$ and $\Delta_K$ to grow at a rate similar to $\omega_\mathrm{ti}$. However, nonlinear saturation ultimately limits growth of $\delta_K$ in atmospheres with $\tilde{\omega}_\mathrm{BV} \gg \omega_\mathrm{ti}$, for reasons discussed in Appendix \ref{app:gMode_Saturation}.}
    \label{fig:Schematic2}
\end{figure}

We can use those frequency solutions to determine the relationships between entropy perturbations and displacement amplitudes. From equation (\ref{eq:TI}) we infer that local entropy contrast $\delta_K$ is related to vertical displacement amplitude $\xi_\parallel$ through
\begin{equation}
    \delta_K 
        = - \left( \frac {\omega} {\omega - i \omega_{\rm ti}} \right) 
                \frac {\xi_\parallel} {\lambda_K}
    \label{eq:deltaK_xiparallel}
    \; \; .
\end{equation}
Also, the entropy difference $\Delta_K$ between a perturbed fluid element and its unperturbed counterpart, is related to $\delta_K$ and $\xi_\parallel$ via
\begin{equation}
    \Delta_K = \left( \frac {i \omega_{\rm ti}} {\omega} \right) \delta_K 
            = - \left( \frac {i \omega_{\rm ti}} {\omega - i \omega_{\rm ti}} \right)
                    \frac {\xi_\parallel} {\lambda_K}
    \; \; .
\end{equation}
Notice that the amplitude of $\Delta_K$ for oscillating perturbations with $\omega_{\rm ti} < \tilde{\omega}_{\rm BV}$ is smaller than the amplitude of $\delta_K$ by a factor similar to $\omega_{\rm ti} / \tilde{\omega}_{\rm BV}$. 

Nothing in the preceding linearized analysis prevents thermally unstable $g$-modes from achieving nonlinear amplitudes and producing precipitation. However, numerical simulations of unmagnetized atmospheres with $\omega_\mathbf{BV} \gg \omega_\mathbf{ti}$ show that the amplitudes of thermally unstable $g$-mode amplitudes saturate before becoming large enough to produce precipitation. Instead of growing indefinitely, the mode amplitudes saturate near $\delta_K \sim \omega_{\mathrm{ti}} / \omega_\mathrm{BV} \sim t_\mathrm{ff} / t_{\rm cool}$, where $t_\mathrm{ff} \equiv (2 r / g)^{1/2}$ is the local freefall time at radius $r = | \mathbf{r}|$ in a potential well with gravitational acceleration $g = | \mathbf{g(r)} |$  \citep[e.g.,][]{McCourt_2012,Wibking_2025}. Appendix \ref{app:gMode_Saturation} discusses how nonlinear saturation arises, and Appendix \ref{app:Freefall} discusses how atmospheric motions can promote precipitation by suppressing buoyancy. But this tutorial's main task is to illuminate how \textit{magnetic forces} enable precipitation to develop in galactic atmospheres that otherwise would not precipitate.

\section{Magnetic Acceleration}
\label{sec:MagAccel}

When the approximations of magnetohydrodynamics are valid, the magnetic acceleration $\mathbf{a_B}$ of an atmospheric fluid element is given by equation (\ref{eq:MagneticAcceleration}), and fluid motions induce changes in the magnetic field according to 
\begin{equation}
    \frac {d \mathbf{B}} {d t}
        = (\mathbf{B} \cdot \nabla) \mathbf{v} - \mathbf{B} (\nabla \cdot \mathbf{v}) 
        \; \; .
    \label{eq:MHD_induction}
\end{equation}
We will assume that the background flow $\mathbf{\tilde{v}}$ is sufficiently uniform for gradients in the displacement field $\boldsymbol{\xi} (\mathbf{r})$ to be much greater than gradients in $\mathbf{\tilde{v}}(\mathbf{r})$. In that case, the magnetic field perturbation resulting from small displacements is
\begin{equation}
    \mathbf{\Delta B} 
        \: = \: (\mathbf{B} \cdot \nabla) \boldsymbol{\xi} 
            - \mathbf{B} (\nabla \cdot \boldsymbol{\xi}) 
        \: = \:  (\mathbf{B} \cdot \nabla) \boldsymbol{\xi} 
            + \mathbf{B} \Delta_\rho 
    \label{eq:DeltaB}
\end{equation}
in which $\Delta_\rho \equiv - \nabla \cdot \boldsymbol{\xi}$ is a compression factor arising from the displacements.\footnote{Equation (\ref{eq:DeltaB}) has introduced a subtle constraint, because it assumes that both density perturbations and magnetic field perturbations start at zero and therefore share the same phase. A more general analysis would include an initial magnetic-field perturbation $\mathbf{\Delta B}(\mathbf{r},0)$ and possibly also a non-zero initial displacement $\boldsymbol{\xi}(\mathbf{r},0)$ consistent with $\Delta_\rho (\mathbf{r},0) = - \nabla \cdot \boldsymbol{\xi}(\mathbf{r},0)$.} 

The first term on the right of equation (\ref{eq:DeltaB}) represents changes resulting from stretching of the magnetic field, while the second one accounts for changes resulting from compression of the field. Displacements that stretch the field without compressing it very much can therefore make the term proportional to $\nabla \boldsymbol{\xi}$ considerably greater than the term proportional to $\Delta_\rho$. However, magnetic field amplification becomes non-linear when $| \nabla \boldsymbol{\xi} |$ approaches unity. The challenge we will eventually face is understanding what then happens to the perturbation.

To simplify the following calculations, we will assume that the unperturbed magnetic field $\mathbf{B}_0$ is uniform. The linearized magnetic field perturbation is then
\begin{equation}
    \mathbf{\Delta B} 
        = ( \mathbf{B}_0 \cdot \nabla) \boldsymbol{\xi} 
                + \mathbf{B}_0 \Delta_\rho 
        \; \; .
\end{equation}
Inserting $\mathbf{B = B_0 + \Delta B}$ into equation (\ref{eq:MagneticAcceleration}) and retaining only terms of linear order gives the magnetic acceleration
\begin{equation}
    \mathbf{a_B} 
        = \frac {(\nabla \times \mathbf{\Delta B} ) \times  
            \mathbf{B_0}} {4 \pi \rho_0}
        = \frac {(\mathbf{B}_0 \cdot \nabla) \mathbf{\Delta B} 
            - \nabla ( \mathbf{B}_0 \cdot \mathbf{\Delta B} )} 
                {4 \pi \rho_0}
    \; \; .
\end{equation}
There is a contribution that depends on stretching of the field and a contribution that depends on compression of the field. We will deal with compression first. 

The compression term in the expression for $\mathbf{\Delta B}$ contributes a magnetic acceleration 
\begin{equation}
    \frac {\mathbf{B}_0 (\mathbf{B}_0 \cdot \nabla) \Delta_\rho 
        - B_0^2 \nabla \Delta_\rho} {4 \pi \rho_0}
        = - v_{\rm A}^2 
            \left[ \nabla \Delta_\rho
                - \mathbf{\hat{B}_0} (\mathbf{\hat{B}_0} \cdot \nabla \Delta_\rho ) \right] 
    \; \; .
\end{equation}
Here, $v_{\rm A}^2 \equiv B_0^2 / 4 \pi \rho_0$ is the square of the Alfv\'en speed $v_{\rm A}$ in the unperturbed medium, which has a magnetic field strength $B_0 = | \mathbf{B}_0 |$, and $\mathbf{\hat{B}_0}$ is a unit vector pointing in the direction of $\mathbf{B}_0$. The term containing $\mathbf{\hat{B}}_0$ cancels the component of $\nabla \Delta_\rho$ aligned with $\mathbf{B}_0$ because the magnetic field does not apply a force in that direction, in a first-order approximation.

The stretching term in the expression for $\mathbf{\Delta B}$ is the one responsible for Alfv\'en waves. To assess its contribution to magnetic acceleration, we will consider a displacement field $\boldsymbol{\xi}$ that is sinusoidal with wavenumber $k_B \equiv \mathbf{k} \cdot \mathbf{\hat{B}}_0$ along the magnetic field, giving us
\begin{equation}
    \frac {(\mathbf{B}_0 \cdot \nabla)^2 \boldsymbol{\xi} 
        - \nabla [ \mathbf{B}_0 \cdot (\mathbf{B}_0 \cdot \nabla) \boldsymbol{\xi} ]} {4 \pi \rho_0}
        = - k_B^2 v_{\rm A}^2 
            \left[ \boldsymbol{\xi}
           - (\mathbf{\hat{B}}_0 \cdot \boldsymbol{\xi} )
                    \mathbf{\hat{B}}_0 \right]
        \; \; .
\end{equation} 
Here again, there is a correction term containing $\mathbf{\hat{B}}_0$, this one canceling the component of $k_{\rm B}^2 v_{\rm A}^2 \boldsymbol{\xi}$ parallel to $\mathbf{B}_0$. 

We will incorporate the effects of magnetic-field stretching into the momentum equation by defining two Alfv\'en-wave frequencies:
\begin{eqnarray}
    \omega_{\rm A,\parallel}^2
        & \equiv & k_B^2 v_{\rm A}^2 
            \left( 1 - \frac {B_{0,\parallel}} {B_0} \right) 
        \\
    \omega_{\rm A,\perp}^2
        & \equiv & k_B^2 v_{\rm A}^2 
            \left( 1 - \frac {B_{0,\perp}} {B_0} \right) 
        \; \; ,
\end{eqnarray}
in which $B_{0,\parallel}$ is the component of $\mathbf{B}_0$ aligned with the displacement $\boldsymbol{\xi}_\parallel$ parallel to $\mathbf{\tilde{g}}$ and $B_{0,\perp}$ is the component
aligned with the displacement $\boldsymbol{\xi}_\perp$ perpendicular to $\mathbf{\tilde{g}}$. The magnetic acceleration response to a displacement field $\boldsymbol{\xi}(\mathbf{r})$ can then be written as 
\begin{equation}
    \mathbf{a_B}
        = - \, \omega_{\rm A,\parallel}^2 \boldsymbol{\xi}_\parallel
            - \omega_{\rm A,\perp}^2 \boldsymbol{\xi}_\perp
            - v_{\rm A}^2 ( \nabla \Delta_\rho )_\mathbf{B}
            \; \; ,
\end{equation}
where $\omega_{\rm A,\parallel}$ is the frequency of an Alfv\'en wave with displacements aligned with gravity, $\omega_{\rm A,\perp}$ is the frequency of an Alfv\'en wave with displacements perpendicular to gravity, and $(\nabla \Delta_\rho)_\mathbf{B} \equiv [ \nabla \Delta_\rho - \mathbf{\hat{B}_0} (\mathbf{\hat{B}_0} \cdot \nabla \Delta_\rho ) ]$ is the component of $\nabla \Delta_\rho$ perpendicular to $\mathbf{B}_0$.

\section{Magnetothermal Drip I}
\label{sec:MTD1}

Adding a uniform magnetic field to a galactic atmosphere introduces a new thermally unstable wave mode, which this tutorial is calling \textit{magnetothermal drip}. An intuitive understanding of that mode benefits from a treatment that is as basic as possible, and so this section will ignore both pressure perturbations and magnetic field compression while considering the effects of a uniform background field perpendicular to gravity. The section that follows (\S \ref{sec:Saturation}) will consider what happens to magnetothermal drip when it becomes nonlinear. After that is a section (\S \ref{sec:PressurePerts}) looking at pressure perturbations. Section \ref{sec:MTD2} then discusses what transpires when the background magnetic field has a vertical component.

Ignoring pressure perturbations and compression in a medium with a uniform horizontal magnetic field reduces equation (\ref{eq:xi_momentum}) to
\begin{equation}
    \ddot{\xi}_\parallel 
        = \tilde{\omega}_{\rm BV}^2 \lambda_K \delta_K 
            - \omega_{\rm A}^2 \xi_\parallel
        \label{eq:ddot_xiparallel}
        \; \; .
\end{equation}
According to this equation, both buoyancy and magnetic tension can accelerate the displacement component $\xi_\parallel$ that is parallel to $\tilde{\mathbf{g}}$. And if buoyancy is negligible, then magnetic tension causes Alfv\'en waves with a horizontal wavenumber $k_\perp$ to oscillate with frequency $\omega_{\rm A} = k_\perp v_{\rm A}$. Assuming that all perturbed quantities are $\propto e^{-i \omega t}$ and making use of equation (\ref{eq:deltaK_xiparallel}) to substitute for $\delta_K$ then gives us
\begin{equation}
        \omega^3 
            -  i \omega_{\rm ti} \omega^2
            - (  \tilde{\omega}_{\rm BV}^2  
                + \omega_{\rm A}^2 ) \omega 
            +  i \omega_{\rm ti} \omega_{\rm A}^2
            = 0
            \; \; .
\end{equation}
This equation's solutions are the frequencies of the simplified system's three normal modes. 

If $\omega_{\rm ti}$ is small compared to $\tilde{\omega}_{\rm BV}$, then the approximate frequencies of the normal modes are
\begin{equation}
    \omega_\pm 
            \approx \, \pm \, ( \tilde{\omega}_{\rm BV}^2 + \omega_{\rm A}^2 )^{1/2}
                \, + \, 
                \frac {i \omega_{\rm ti}} {2}
                \left( \frac {\tilde{\omega}_{\rm BV}^2} 
                        {\tilde{\omega}_{\rm BV}^2 + \omega_{\rm A}^2 } \right)
\end{equation}
\begin{equation}
    \omega_0 \approx i \omega_{\rm ti}
            \left( \frac {\omega_{\rm A}^2} 
                        {\tilde{\omega}_{\rm BV}^2 
                            + \omega_{\rm A}^2} \right)
        \; \; .
\end{equation} 
All three modes are thermally unstable for $\omega_{\rm ti} > 0$. The modes with frequencies $\omega_\pm$ are magneto-gravity waves, similar to pure internal gravity waves but with frequencies enhanced by magnetic tension \citep[e.g.,][]{SteinLeibacher_1974ARA&A..12..407S}. The third mode, with frequency $\omega_0$, is what we are calling magnetothermal drip. 

\begin{figure}[t]
    \centering
    \includegraphics[width=0.99\linewidth]{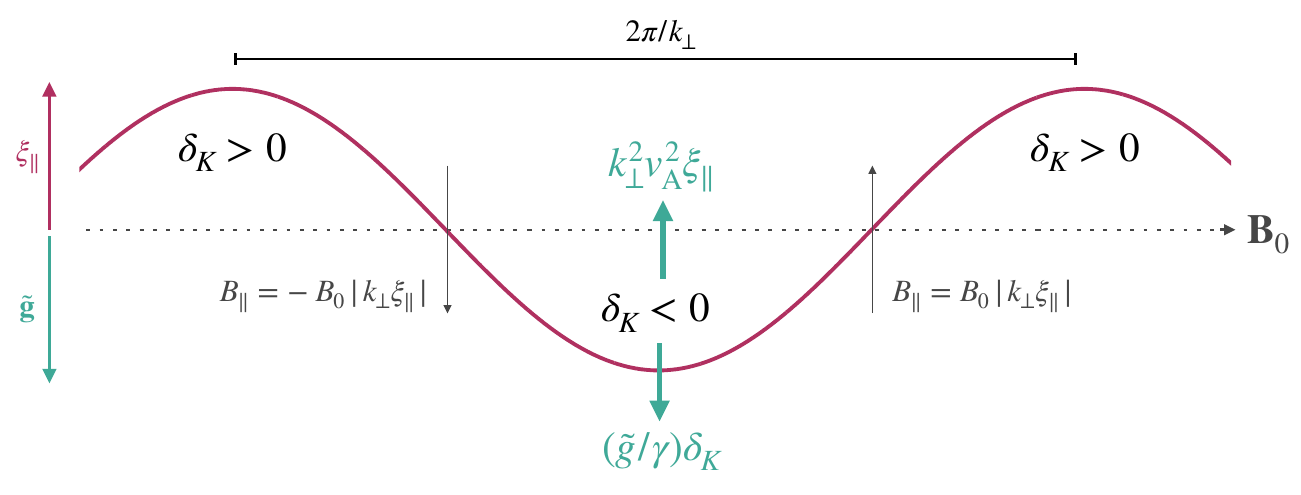}
    \caption{Schematic illustration of linear magnetothermal drip in a stratified atmosphere with a uniform horizontal background magnetic field ($\mathbf{B}_0$, grey dotted line). Vertical displacement ($\xi_\parallel$, solid red line), entropy contrast ($\delta_K$), and vertical magnetic field strength ($B_\parallel$) all steadily grow on a thermal-instability timescale similar to the cooling time when magnetic acceleration ($k_\perp^2 v_{\rm A}^2 \xi_\parallel$) nearly offsets buoyant acceleration ($\tilde{g} \delta_K / \gamma$).}
    \label{fig:MHD_Drip_linear}
\end{figure}

Magnetothermal drip steadily grows because the buoyancy and magnetic tension terms in equation (\ref{eq:ddot_xiparallel}) nearly offset each other when $\omega = \omega_0$ and $\omega_{\rm ti} \ll \tilde{\omega}_{\rm BV}$. A low-entropy perturbation therefore slowly descends as its entropy decreases, and its descent strengthens the vertical component of the magnetic field supporting it (see Figure \ref{fig:MHD_Drip_linear}). A sinusoidal magnetothermal drip mode therefore resembles an Alfv\'en wave with an amplitude that exponentially increases without oscillating. The mode grows on a timescale similar to $t_{\rm cool}$ as long as its wavenumber exceeds a transitional wavenumber at which $\omega_{\rm A} = \tilde{\omega}_{\rm BV}$:
\begin{equation}
    k_0 \: \equiv \: \frac {\tilde{\omega}_{\rm BV}} {v_{\rm A}}
        \: = \: \left( \frac {|\mathbf{\tilde{g}}|} 
                        {\gamma \lambda_K v_{\rm A}^2} \right)^{1/2}
        \: = \: \left( \frac {\beta_0} 
                    {2 \gamma \lambda_K \lambda_P} \right)^{1/2}
        \; \; .
\end{equation}
On the right, $\beta_0 \equiv 8 \pi P_0 / B_0^2$ is the background ratio of thermal pressure to magnetic pressure, and $\lambda_P \equiv P_0 / \rho_0 |\mathbf{\tilde{g}}|$ is the atmosphere's pressure scale height. The characteristic wavenumber of magnetothermal drip in a galactic atmosphere with $\lambda_K \sim \lambda_P$ therefore scales as $k_0 \sim \beta_0^{1/2} \lambda_P^{-1}$.

While the name ``magnetothermal drip" may be new, the mode itself is not. It is one of the normal modes that \citet{Loewenstein_1990} found in his linearized analysis of magnetothermal instability. It is also present in the numerical simulations of \citet{Ji_2018}, which demonstrated that magnetic fields could promote development of nonlinear density contrasts via thermal instability in atmospheres that would otherwise not precipitate. However, to understand how magnetothermal drip manages to culminate in precipitation, we need to extend our stability analysis into the nonlinear regime.

According to the linearized analysis, the vertical magnetic field of a magnetothermal drip mode has a magnitude $|B_\parallel| = B_0 | k_\perp  \xi_\parallel |$, and the forces it applies produce a vertical acceleration
\begin{equation}
    a_{\mathbf{B},\parallel} 
        = - k_\perp^2 v_{\rm A}^2 \xi_\parallel
        \: \propto \frac {| B_\parallel| } {B_0} 
        \; \; .
\end{equation}
However, amplification of the vertical field becomes nonlinear when $| k_\perp \xi_\parallel | \sim 1$. At that point, the entropy contrast of a magnetothermal drip mode is 
\begin{equation}
    \delta_K 
        \: = \: \left( \frac {\omega_{\rm A}^2 - \omega_0^2} 
                        {\tilde{\omega}_{\rm BV}^2} \right) 
                    \frac {\xi_\parallel} {\lambda_K}
        \: \approx \: \left( \frac {k_\perp} {k_0} \right)^2  
                    \frac {\xi_\parallel} {\lambda_K}
        \label{eq:deltaK_drip}
        \; \; .
\end{equation}
Drip modes with $k_\perp \sim k_0$ therefore reach the nonlinear regime with entropy contrast 
\begin{equation}
    \delta_K 
        \: \sim \: \frac {1} {k_0 \lambda_K}  
        \: = \:  \left( \frac {2 \gamma \lambda_P} {\beta_0 \lambda_K} \right)^{1/2}
        \; \; .
\end{equation}
Consequently, $\delta_K$ is still small when magnetic-field amplification becomes nonlinear in an atmosphere with $\beta_0 \gg 1$ and $\lambda_K \sim \lambda_P$.

Notice also that magnetothermal drip modes with $k_\perp \sim k_0$ are the first modes to undergo nonlinear magnetic field amplification: Longer wavelength drip modes $(k_\perp \ll k_0)$ grow more slowly (by a factor $\sim k_\perp^2 / k_0^2$). And shorter wavelength drip modes $(k_\perp \gg k_0)$ with similar entropy contrast have less magnetic-field amplification (because $B_\parallel \propto  k_\perp  \xi_\parallel \propto \delta_K / k_\perp$).

Magnetothermal drip can proceed toward precipitation as long as the vertical magnetic tension supporting the overdense gas continues to strengthen in proportion to $\xi_\parallel$ and $\delta_K$. But when does the perturbation's entropy (and density) contrast become nonlinear? According to the linearized analysis resulting in equation (\ref{eq:deltaK_drip}), the amplitude $\delta_K$ of a drip mode with wavenumber $\sim k_0$ becomes nonlinear as its vertical displacement approaches the atmosphere's entropy scale height (i.e., $\xi_\parallel \sim -\lambda_K$). At that point, the strength of the magnetic field supporting the denser gas is 
\begin{equation}
    B_\parallel 
        \: \approx \: B_0 (k_0 \lambda_K) 
        \: \sim \: \beta_0^{1/2} B_0 
        \; \; .
\end{equation}
In other words, the local magnetic pressure $(B^2 / 8 \pi)$ associated with amplification of $B_\parallel$ has become comparable to the atmosphere's thermal pressure. Large magnetic field amplification factors local to condensed gas are therefore another signature of nonlinear magnetothermal drip in atmospheres with $\beta_0 \gg 1$.

However, numerical simulations demonstrate that nonlinear magnetothermal drip modes do not always attain nonlinear density contrasts. For example, \citet{Ji_2018} found that magnetothermal drip saturates with
\begin{equation}
    \frac {\delta \rho} {\rho} 
        \sim 3 \beta_0^{-1/2} 
        \left( \frac {t_{\rm cool}} {t_{\rm ff}} \right)^{-1}
\end{equation}
in which $t_{\rm cool} / t_{\rm ff} \sim \tilde{\omega}_{\rm BV} / \omega_{\rm ti}$ is the background ratio of cooling time to freefall time. We would therefore like to know \textit{how} magnetothermal drip saturates and \textit{why} saturation depends on both $\beta_0$ and $t_{\rm cool} / t_{\rm ff}$.

\section{Nonlinear Saturation}
\label{sec:Saturation}

To understand nonlinear saturation of magnetothermal drip, we need to focus on the induction equation for the vertical magnetic field, because that is where nonlinearity first arises. The vertical part of equation (\ref{eq:MHD_induction}) can be written as
\begin{equation}
    \frac {1} {B_\parallel} \frac {d B_\parallel} {dt}
        = \frac {( \mathbf{B}_\perp \cdot \nabla_\perp) 
                    \dot{\xi}_\parallel}
                {B_\parallel}
            - \nabla_\perp \cdot \boldsymbol{\dot{\xi}}_\perp 
            \label{eq:dBperp_dt}
            \; \; .
\end{equation}
Here, $\mathbf{B}_\perp$ is the component of magnetic field perpendicular to $\mathbf{\tilde{g}}$, $\boldsymbol{\xi}_\perp$ is the displacement component  perpendicular to $\mathbf{\tilde{g}}$, and $\nabla_\perp$ is a differential operator that applies partial derivatives in directions perpendicular to $\mathbf{\tilde{g}}$. 

So far, we have been ignoring perpendicular displacements, but now we need to consider them because they can limit growth of $B_\parallel$. The first term on the right of equation (\ref{eq:dBperp_dt}) is the amplification term we have already discussed, and its magnitude for a magnetothermal drip mode is similar to $\omega_{\rm ti}$. The second term on the right can therefore offset magnetic field amplification if $\boldsymbol{\xi}_\perp$ is diverging and $\nabla_\perp \cdot \boldsymbol{\dot{\xi}}_\perp$ is also similar to $\omega_{\rm ti}$.

Now consider the low-entropy gas in a magnetothermal drip perturbation that is descending on a cooling timescale. As it descends, it loses entropy and enters layers of greater pressure. Both of those effects can increase the perturbation's density, further amplifying $B_\parallel$, but there is a countervailing effect. As magnetothermal drip amplifies the vertical field, it produces horizontal gradients in magnetic pressure that are proportional to $\xi_\parallel^2$. Those gradients generate horizontal accelerations that can push plasma away from the extrema in $\delta_K$, thereby limiting amplification of $B_\parallel$.

The horizontal accelerations stemming from vertical field amplification are second-order in the perturbation amplitude, because the magnetic pressure gradient associated with the vertical field is proportional to $B_\parallel^2$. As long as the linear approximation $B_\parallel \approx (\mathbf{B}_0 \cdot \nabla_\perp) \xi_\parallel$ remains valid, the horizontal acceleration produced by gradients in the vertical field is
\begin{equation}
    \boldsymbol{\ddot{\xi}}_\perp =
        - \frac {\nabla_\perp B_\parallel^2} {8 \pi \rho_0}  
        \: \sim \: \frac {\omega_{\rm A}^2 k_\perp \xi_\parallel^2} {2} 
        \; \; .
\end{equation}
To obtain a cumulative horizontal displacement, we can integrate $\xi_\parallel^2 \propto e^{2 \omega_{\rm ti} t}$ twice with respect to time, giving
\begin{equation}
    | k_\perp \xi_\perp | 
        \sim \frac {\omega_{\rm A}^2} {8 \omega_{\rm ti}^2} 
            | k_\perp \xi_\parallel |^2
    \; \; .
\end{equation}
The product $k_\perp \xi_\perp$ therefore becomes nonlinear for
\begin{equation}
    | k_\perp \xi_\parallel | \sim 3 \frac {\omega_{\rm ti}} {\omega_{\rm A}}
    \; \; .
\end{equation}
However, horizontal displacements strongly interfere with each other as $|k_\perp \xi_\perp|$ approaches unity, meaning that the linearized field amplification relation $| B_\parallel | \approx B_0 | k_\perp \xi_\parallel |$ cannot hold beyond that point, and magnetothermal drip presumably saturates.

The saturation criterion that has emerged aligns well with the simulation results of \citet{Ji_2018}. Modes with $k_\perp \sim k_0$ have $\omega_{\rm A} \sim \tilde{\omega}_{\rm BV}$ and $\omega_{\rm ti} / \omega_{\rm A} \sim (t_{\rm cool} / t_{\rm ff})^{-1}$. Consequently, the amplitude of magnetothermal drip when horizontal displacements saturate in an atmosphere with $\lambda_K \sim \lambda_P$ should be
\begin{equation}
    \delta_K \sim \frac {\xi_\parallel} {\lambda_K}
        \sim 3 \beta_0^{-1/2} 
                \left( \frac {t_{\rm cool}} {t_{\rm ff}} \right)^{-1}
    \; \; .
\end{equation}
This line of argument does not \textit{prove} that horizontal magnetic pressure gradients are solely responsible for the saturation observed in simulations, but it does provide a plausible explanation for why the scaling of saturated magnetothermal drip depends on both $\beta_0$ and $t_{\rm cool} / t_{\rm ff}$.

\begin{figure}[t]
    \centering
    \includegraphics[width=0.99\linewidth]{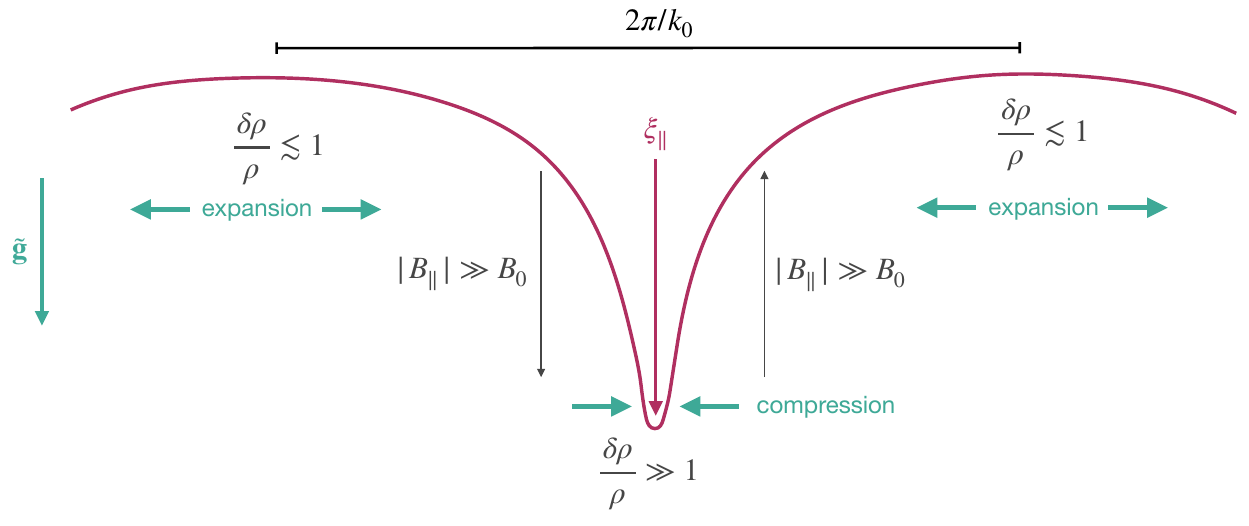}
    \caption{Schematic illustration of nonlinear magnetothermal drip in a stratified atmosphere with a uniform horizontal magnetic field of strength $B_0$. A solid red line shows vertical displacement ($\xi_\parallel$), relative to the unperturbed background state. Growth and descent of a nonlinear density perturbation ($\delta \rho / \rho$) has pulled the magnetic field lines threading it downward, and simultaneous amplification of the vertical magnetic field ($B_\parallel$) has kept it strong enough to counteract negative buoyancy. Low-entropy gas in the perturbation has become more compressed because of both entropy losses and descent through higher-pressure layers. Meanwhile, higher entropy gas has become less compressed because it has ascended to lower-pressure layers. The magnetic field lines and displacements of an originally sinusoidal perturbation consequently become asymmetric.
    }
    \label{fig:MHD_Drip_nonlinear}
\end{figure}

Another feature of nonlinear magnetothermal drip is asymmetry between descending low-entropy perturbations and ascending high-entropy perturbations. As mentioned earlier, a low-entropy perturbation becomes more highly compressed as it descends. If it can cool fast enough, then horizontal thermal-pressure gradients will continue to compress it until the local value of $\beta$ approaches unity. Meanwhile, high-entropy perturbations expand as they ascend, and horizontal decompression hinders growth of the vertical field component. We therefore expect a magnetothermal drip mode that starts out nearly sinusoidal, when its amplitude is small, to become asymmetrically pinched at lower altitudes and distended at higher altitudes (see Figure \ref{fig:MHD_Drip_nonlinear}). Such magnetic-field morphologies are indeed observed in numerical simulations of nonlinear magnetothermal drip \citep{Wibking_2025MNRAS.544.2577W}; section \ref{sec:MHDPrecip} presents some examples.

\section{Pressure Perturbations}
\label{sec:PressurePerts}

To prepare for the next section's analysis of magnetothermal instability in atmospheres with vertical magnetic fields, we first need to incorporate pressure perturbations. This section will look at their effects on thermal instability in \textit{unmagnetized} atmospheres and will develop formalism to be used in the section that follows. It will demonstrate a feature of pressure perturbations that is already well known \citep[e.g.,][]{Binney_2009}: They slow buoyant oscillations somewhat by coupling vertical displacements with horizontal displacements. 

An important constraint on those displacements comes from the continuity equation for gas mass:
\begin{equation}
    \frac {d \ln \rho} {dt} = - \nabla \cdot \mathbf{v}
    \; \; .
\end{equation}
Subtracting the continuity equation for the unperturbed flow from the continuity equation for the perturbed flow gives a continuity equation for density perturbations:
\begin{equation}
    \dot{\Delta}_\rho \equiv \frac {d \ln \rho} {dt} - \frac {d \ln \rho_0} {dt} 
        = - \nabla  \cdot \boldsymbol{\dot{\xi}}
        \; \; .
\end{equation}
As before, $\Delta_\rho = - \nabla \cdot \boldsymbol{\xi}$ is the net compression applied by the displacement field.

To describe evolution of the pressure perturbations, let $\Delta_P$ represent the pressure difference between a perturbed fluid element and its unperturbed counterpart:
\begin{equation}
    \Delta_P (\mathbf{r}_0,t) 
        \equiv \ln \left[ \frac {P(\mathbf{r}_0,t)} {P_0(\mathbf{r}_0,t)} \right]
    \; \; . 
\end{equation}
From the equation of state we get the relationship between $\Delta_P$, $\Delta_K$, and $\Delta_\rho$:
\begin{equation}
    \Delta_P = \Delta_K + \gamma \Delta_\rho
    \; \; .
\end{equation}
The pressure $P(r)$ of a perturbed fluid element at location $\mathbf{r}$ is then given by
\begin{equation}
    \ln P (\mathbf{r}) 
        = \ln P_0 (\mathbf{r} - \boldsymbol{\xi}) + \Delta_K + \gamma \Delta_\rho 
        \; \; ,
\end{equation}
where $P_0 (\mathbf{r} - \boldsymbol{\xi})$ is the pressure that the same fluid element \textit{would} have if it were unperturbed. If we disregard corrections that are second order in $\xi_\parallel$, the local pressure contrast factor can be expressed as
\begin{equation}
    \delta_P ( \mathbf{r} ) = \ln P (\mathbf{r}) - \ln P_0(\mathbf{r})
        = \gamma \Delta_\rho + \Delta_K + \frac {\xi_\parallel} {\lambda_P}
        \label{eq:delta_P}
        \; \; ,
\end{equation}
where $\lambda_P \equiv | \nabla \ln P_0 |^{-1} = c_{\rm s}^2 / \gamma |\mathbf{\tilde{g}}|$ is the local scale height for a pressure gradient that points in the same direction as $\mathbf{\tilde{g}}$. Equation (\ref{eq:delta_P}) will be useful because it neatly separates the contributors to $\delta_P$ into a part coming from compression $(\gamma \Delta_\rho)$, a part coming from entropy change $(\Delta_K)$, and a part coming from vertical displacement alone $(\xi_\parallel/ \lambda_P)$.

If magnetic forces are negligible, including pressure perturbations in the vertical part of the momentum equation gives the following expression for the oscillation frequency: 
\begin{equation}
    \omega^2 
        = \left( \frac {\omega} {\omega - i \omega_{\rm ti}} \right) \tilde{\omega}_{\rm BV}^2 
            -  \tilde{g} (\gamma - 1 ) \frac {\delta_P} {\gamma \xi_\parallel}
            + c_{\rm s}^2 \left( \frac {1} {\xi_\parallel} 
                                \frac {\partial} {\partial \tilde{r}} 
                                \frac {\delta_P} {\gamma} \right)
        \; \; ,
        \label{eq:omega2}
\end{equation}
where $\tilde{g} = |\mathbf{\tilde{g}}|$ is the magnitude of effective gravitational acceleration and $\tilde{r}$ is a spatial coordinate that increases in the direction opposite to $\boldsymbol{\tilde{g}}$. The last term in this equation requires careful treatment, as it plays two complementary roles: It causes the frequencies of gravity-wave modes to be smaller than $\tilde{\omega}_{\rm BV}$, and it also accounts for acoustic modes with $\omega > \tilde{\omega}_{\rm BV}$.

We will begin our analysis of that term by assuming that all perturbed quantities vary sinusoidally in the horizontal direction, with horizontal wavenumber $k_\perp$. Compression can then be expressed in terms of vertical $(\xi_\parallel)$ and horizontal $(\xi_\perp)$ displacements via
\begin{equation}
    \Delta_\rho = - \nabla \cdot \boldsymbol{\xi} 
        = - i k_\perp \xi_\perp - \frac {\partial \xi_\parallel} {\partial \tilde{r}}
    \; \; . 
\end{equation}
At this point, it is tempting to assume that $\xi_\parallel$ is a sinusoidal function of $\tilde{r}$, but that isn't necessarily a valid assumption, because the background state depends on $\tilde{r}$. Instead, we will focus on the horizontal part of the momentum equation and derive an expression that relates $\xi_\perp$ to pressure perturbations.

Without magnetic forces, pressure perturbations are the only drivers of horizontal displacement $\boldsymbol{\xi}_\perp$, and the horizontal part of the momentum equation reduces to
\begin{equation}
     \omega^2 \xi_\perp 
        = i  k_\perp c_{\rm s}^2 \frac {\delta_P} {\gamma} 
        \label{eq:deltaP_xiperp}
        \; \; .
\end{equation}
This result implies that the pressure perturbation amplitude $\delta_P$ is much smaller than the product $k_\perp \xi_\perp$ for modes that oscillate slowly compared to horizontal sound waves with a similar wavenumber. While that condition holds, displacement speeds are relatively slow $(\omega \xi_\perp \ll c_{\rm s})$ and the magnitude of $\delta_P$ is small compared to $|k_\perp \xi_\perp|$. Furthermore, using equation (\ref{eq:delta_P}) to substitute for $\delta_P / \gamma$, writing $\Delta_\rho$ in terms of displacement amplitudes, and rearranging things yields
\begin{eqnarray}
    \xi_\perp & = & 
        \frac {i k_\perp c_{\rm s}^2} {k_\perp^2 c_{\rm s}^2 - \omega^2 } 
            \left( \frac {\partial \xi_\parallel} {\partial \tilde{r}} 
                - \frac {\Delta_K} {\gamma} 
                - \frac {\xi_\parallel} {\gamma \lambda_P} \right)
        \label{eq:xi_perp}
                    \\
    \frac {\delta_P} {\gamma} & = & 
        \frac {\omega^2} {k_\perp^2 c_{\rm s}^2 - \omega^2 } 
                \left( \frac {\partial \xi_\parallel} {\partial \tilde{r}} 
                - \frac {\Delta_K} {\gamma} 
                - \frac {\xi_\parallel} {\gamma \lambda_P} \right)
        \label{eq:deltaP_gamma}
                    \\
    \Delta_\rho & = & 
        \frac {\omega^2} {k_\perp^2 c_{\rm s}^2 - \omega^2 } 
                \frac {\partial \xi_\parallel} {\partial \tilde{r}} 
                    - \frac {k_\perp^2 c_{\rm s}^2} {k_\perp^2 c_{\rm s}^2 - \omega^2 } 
                        \left( \frac {\Delta_K} {\gamma} 
                    + \frac {\xi_\parallel} {\gamma \lambda_P} \right)
                    \; \; .
\end{eqnarray}
We will see that results like these are helpful tools for interpreting the effects of pressure perturbations. 

For example, consider the limit of oscillations that are slow ($\omega^2 \ll k_\perp^2 c_{\rm s}^2$) and nearly adiabatic $(\Delta_K \ll \partial \xi_\parallel / \partial \tilde{r})$, with a short vertical wavelength ($\partial \xi_\parallel / \partial \tilde{r} \gg \xi_\parallel / \lambda_P$ and $\partial^2 \xi_\parallel / \partial \tilde{r}^2 \gg \tilde{g} / c_{\rm s}^2$). In that case, we can combine equations (\ref{eq:deltaP_xiperp}) and (\ref{eq:xi_perp}) to obtain 
\begin{equation}
    \frac {\delta_P} {\gamma} =  \frac {\omega^2} {k_\perp^2 c_{\rm s}^2}
                \frac {\partial \xi_\parallel} {\partial \tilde{r}} 
                    \; \; .
\end{equation}
Substitution for $\delta_P / \gamma$ in equation (\ref{eq:omega2}) then gives
\begin{equation}
    \omega^2 
        = \frac {k_\perp^2} {k^2} \left( \frac {\omega} {\omega - i \omega_{\rm ti}} \right) \tilde{\omega}_{\rm BV}^2 
       \; \; \; \; , \; \; \; \; 
    k^2 = k_\perp^2 - \frac {1} {\xi_\parallel} 
                                \frac {\partial^2 \xi_\parallel} {\partial \tilde{r}^2} 
            \; \; ,
\end{equation}
in which the quantity $k^2$ is equivalent to the square of the total wavenumber for a perturbation that is sinusoidal in $\tilde{r}$. The system's normal modes in this limit therefore have frequencies
\begin{equation}
    \omega_\pm 
        = \pm \, \omega_{\rm buoy} \left( 1 - \frac {\omega_{\rm ti}^2} {4 \omega_{\rm buoy}^2} \right)^{1/2} 
            + \frac {i \omega_{\rm ti}} {2}
            \; \; ,
\end{equation}
where $\omega_{\rm buoy}^2 \equiv (k_\perp / k)^2 \, \tilde{\omega}_{\rm BV}^2$ specifies the frequency $\omega_{\rm buoy}$ of an adiabatic internal gravity wave with horizontal wavenumber $k_\perp$.

Interpretation of the $k_\perp/k$ factor that has appeared is simple: Pressure perturbations slow the frequency $\omega_{\rm buoy}$ of an internal gravity wave by the ratio $k_\perp / k$ because they produce significant horizontal displacements when $k_\perp^2$ is significantly less than $k^2$. Horizontal motions associated with a certain vertical displacement amplitude are then more pronounced, but the buoyancy forces remain the same.  The additional inertia of horizontal motion is what reduces the accelerations driven by buoyancy.

Thermally unstable modes with longer wavelengths or greater frequencies are a bit more complicated. Appendix \ref{app:AcousticStability} discusses them for the sake of completeness, and also to maintain contact with more general treatments of wave modes in stratified terrestrial and stellar atmospheres. However, the next section will not worry about thermally unstable modes with long wavelengths or rapid frequencies.

\section{Magnetothermal Drip II}
\label{sec:MTD2}

Vertical magnetic fields enter the thermal instability picture when they are strong enough to resist horizontal displacements. To demonstrate how that happens, we will start with the vertical part of equation (\ref{eq:xi_momentum}), including both pressure perturbations and magnetic acceleration: 
\begin{equation}
    \ddot{\xi}_\parallel
        \: = \: \, \tilde{\omega}_{\rm BV}^2 \lambda_K \delta_K
            \: + \: \tilde{g} (\gamma - 1) 
                    \frac {\delta_P} {\gamma} 
                   \: - \: c_{\rm s}^2  
                \frac {\partial} {\partial \tilde{r}}  
                \frac {\delta_P} {\gamma}
            \: - \: \omega_{\rm A,\parallel}^2 \xi_\parallel
            \: - \: v_{\rm A}^2 
                 \left( 1 - \frac {B_{0,\parallel}} {B_0} \right) \, 
                \frac {\partial} {\partial \tilde{r}}  \Delta_\rho
    \label{eq:ddot_xiparallel_MHD}
    \; \; .
\end{equation}
If the background field is purely vertical, then the terms containing $\omega_{\rm A,\parallel}^2$ and $v_{\rm A}^2$ vanish. The effects of vertical magnetic fields on vertical displacement must therefore come entirely from the vertical gradients of thermal pressure perturbations (i.e. $\partial \delta_P / \partial \tilde{r}$).

To simplify the following calculations, we will assume that $c_{\rm s}^2 \gg v_{\rm A}^2$, so that the magnetic compression term containing $v_{\rm A}^2$ can be ignored because it is small compared to the thermal compression term containing $c_{\rm s}^2$. From equation (\ref{eq:ddot_xiparallel_MHD}) we can then obtain an expression similar to equation (\ref{eq:omega2}), using equations (\ref{eq:deltaK_xiparallel}) and (\ref{eq:deltaP_gamma}):
\begin{equation}
    \omega^2 
        = \left( \frac {\omega} 
                         {\omega - i \omega_{\rm ti}} \right) \tilde{\omega}_{\rm BV}^2 
            -  \tilde{g} (\gamma - 1 ) 
                \left( \frac {\Delta_\rho} {\xi_\parallel}
                    + \frac {\Delta_K} {\gamma \xi_\parallel}
                    + \frac {1} {\gamma \lambda_P} \right)
        + \omega_{\rm A,\parallel}^2
        + c_{\rm s}^2 
            \left[ \frac {1} {\xi_\parallel} 
                   \frac {\partial} {\partial \tilde{r}}
                   \left( \Delta_\rho 
                    + \frac {\delta_K} {\gamma} 
                    + \frac {\xi_\parallel} {\gamma \lambda_P} \right)
                                 \right]
        \; \; .
        \label{eq:omega2_MHD}     
\end{equation}
The horizontal part of equation (\ref{eq:xi_momentum}) gives us another equation relating $\omega^2$ and $\Delta_\rho$, similar to equation (\ref{eq:deltaP_xiperp}):
\begin{equation}
    ( \omega^2 - \omega_{\rm A,\perp}^2  ) \xi_\perp
        \: = \: i k_\perp c_{\rm s}^2 
            \frac {\delta P} {\gamma} 
        \: = \: i k_\perp c_{\rm s}^2 
            \left( \Delta_\rho + \frac {\Delta_K} {\gamma}
                + \frac {\xi_\parallel}
                        {\gamma \lambda_P} \right)
        \label{eq:xiperp_B}
        \; \; .
\end{equation}
Writing $\Delta_\rho$ in terms of displacements and solving for $\xi_\perp$ yields
\begin{equation}
    \xi_\perp =  
        \frac {i k_\perp c_{\rm s}^2} 
            {k_\perp^2 c_{\rm s}^2 
                + \omega_{\rm A,\perp}^2  - \omega^2 } 
            \left(
            \frac {\partial \xi_\parallel} {\partial \tilde{r}} 
        - \frac {\Delta_K} {\gamma} 
                - \frac {\xi_\parallel} {\gamma \lambda_P} \right)
                \; \; .
\end{equation}
We can then put that result back into equation (\ref{eq:xiperp_B}) and solve for $\delta_P$ to get
\begin{equation}
    \frac {\delta_P} {\gamma} = 
        \frac {\omega^2 - \omega_{\rm A,\perp}^2} 
            {k_\perp^2 c_{\rm s}^2 
                + \omega_{\rm A,\perp}^2  - \omega^2 }  
        \left( \frac {\partial \xi_\parallel} 
                      {\partial \tilde{r}} 
                + \frac {\Delta_K} {\gamma} 
                    + \frac {\xi_\parallel} {\gamma \lambda_P} \right)
        \label{eq:Compression}
            \; \; .
\end{equation}
This last equation conveniently expresses how pressure perturbations ($\delta_P$) depend on vertical displacement ($\xi_\parallel$) and entropy change $(\Delta_K)$. 

\begin{figure}[t]
    \centering
    \includegraphics[width=0.99\linewidth]{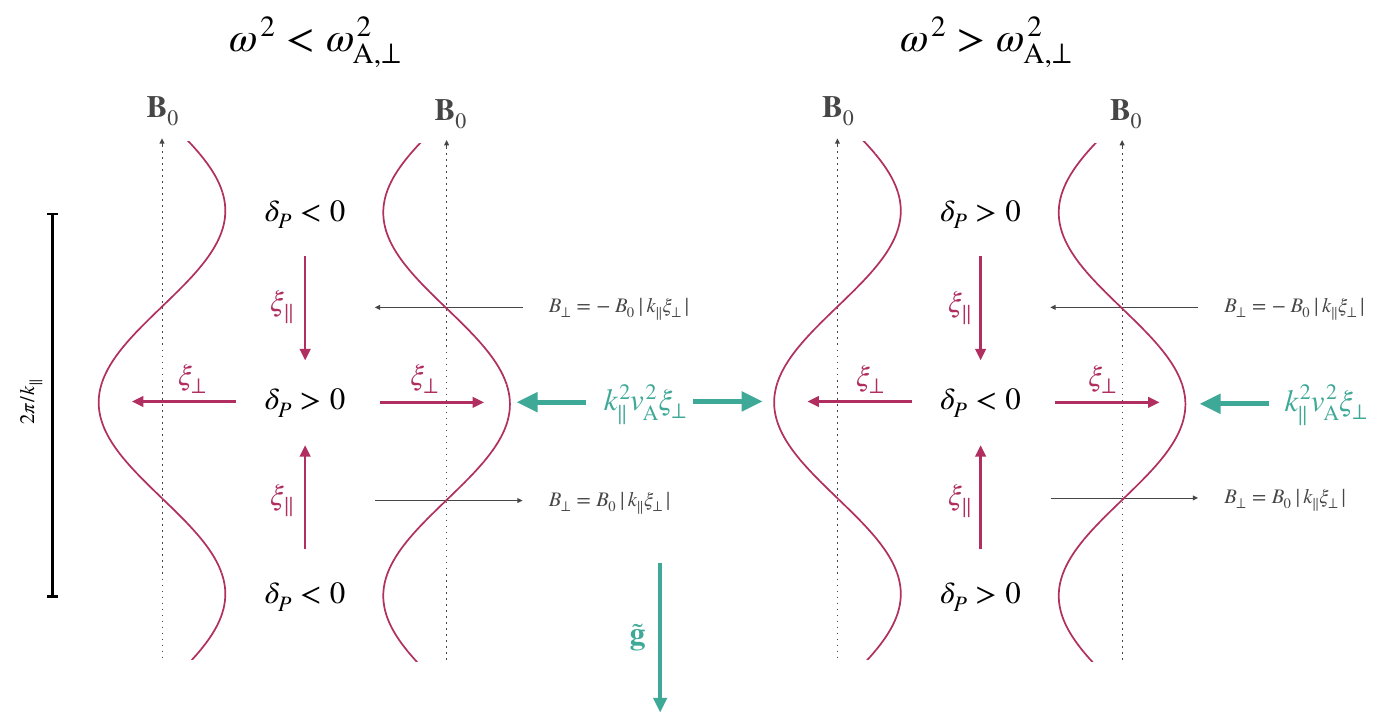}
    \caption{Schematic illustration of perturbations in a stratified atmosphere with a uniform vertical magnetic field of strength $B_0$ (dotted black lines). Solid red lines show horizontal displacement ($\xi_\perp$), relative to the unperturbed background state. Smaller red arrows show the pattern of horizontal and vertical displacements around the central pressure perturbation. Magnetic acceleration of magnitude $k_\parallel^2 v_{\rm A}^2 \xi_\perp$ is resisting the diverging horizontal flow. On the left side, showing a case with $\omega^2 < \omega_{\rm A,\perp}^2$, the central pressure perturbation is positive, because the vertical field successfully resists horizontal divergence. The pressure perturbation therefore inhibits descent of lower-entropy gas lying above it and enables its entropy contrast to grow. On the right side, showing a faster mode with $\omega^2 > \omega_{\rm A,\perp}^2$, horizontal divergence causes the central pressure perturbation to be negative, and it does not resist descent of the gas lying above it.
    }
    \label{fig:MHD_Drip_Vertical}
\end{figure}

To appreciate how vertical magnetic fields affect thermal pressure gradients, according to this result for $\delta_P$, we will temporarily ignore the terms containing $\Delta_K$ and $\xi_\parallel / \lambda_P$, and we will consider frequencies that are slow compared to the acoustic frequency $k_\perp c_{\rm s}$. In that limit, equation (\ref{eq:Compression}) reduces to
\begin{equation}
    \Delta_\rho 
        \: = \: \frac {\delta_P} {\gamma}
        \: = \: \frac {\omega^2 - \omega_{\rm A,\perp}^2} 
              {k_\perp^2 c_{\rm s}^2} 
                \frac {\partial \xi_\parallel} 
                      {\partial \tilde{r}} 
            \; \; .
\end{equation}
It implies that both compression and the local pressure perturbation are directly proportional to the gradient of vertical displacement. However, the \textit{sign} of those perturbations depends on the sign of $\omega^2 - \omega_{\rm A,\perp}^2$. A dependence on vertical magnetic field strength has arisen (via $\omega_{\rm A,\perp}^2$) because vertical magnetic fields resist horizontal plasma motions, and therefore alter the \textit{qualitative} behavior of modes with $\omega^2 \lesssim \omega_{\rm A,\perp}^2$. Without magnetic fields, increased compression $(\Delta_\rho > 0)$ drives vertical divergence ($\partial \xi_\parallel / \partial \tilde{r} > 0)$, and decreased compression $(\Delta_\rho < 0)$ allows vertical convergence ($\partial \xi_\parallel / \partial \tilde{r} < 0)$. However, among magnetized modes with $\omega^2 < \omega_{\rm A}^2$, vertical convergence \textit{increases} compression and produces a pressure increase (see Figure \ref{fig:MHD_Drip_Vertical}), because vertical fields are able to \textit{restrict divergence of horizontal displacements.}

Inserting the approximation for compression into equation (\ref{eq:omega2_MHD}) and dropping the terms proportional to $\Delta_K$, $\tilde{g} / \lambda_P$, and $\tilde{g}/\xi_\parallel$ then gives us
\begin{equation}
    \omega^3 - i \omega_{\rm ti} \omega^2 
        - ( \omega_{\rm buoy}^2 + \omega_{\rm A}^2 ) \, \omega 
        + i \omega_{\rm ti} \omega_{\rm A}^2
            = 0
\end{equation}
\begin{equation}
    \omega_{\rm A}^2
        =  \left( \frac {k_\perp^2} {k^2} \right)
            \omega_{{\rm A},\parallel}^2
            + \left( 1- \frac {k_\perp^2} {k^2} \right)
                \omega_{{\rm A},\perp}^2
        \; \; .
\end{equation}
In other words, vertical magnetic fields alter both the oscillation frequencies of the gravity-wave modes and the growth rate of magnetothermal drip, because the approximate frequency solutions have become
\begin{equation}
    \omega_\pm 
            \approx \, \pm \, ( \omega_{\rm buoy} + \omega_{\rm A} )^2
                \, + \, 
                \frac {i \omega_{\rm ti}} {2}
                \left( \frac {\omega_{\rm buoy}^2} 
                    {\omega_{\rm buoy}^2 + \omega_{\rm A}^2} \right)
\end{equation}
\begin{equation}
    \omega_0 \approx i \omega_{\rm ti}
            \left( \frac {\omega_{\rm A}^2} 
                        {\omega_{\rm buoy}^2 + \omega_{\rm A}^2}  \right)
        \; \; .
\end{equation}
The form of these equations is essentially identical to the result for a purely horizontal background field, derived much more simply in section \ref{sec:MTD1}, except that $\omega_{\rm buoy} = (k_\perp/k) \tilde{\omega}_{\rm BV}$ has replaced $\tilde{\omega}_{\rm BV}$ and $\omega_{\rm A}$ now accounts for background magnetic fields with a vertical component.

Importantly, and perhaps counterintuitively, thermally unstable growth of short-wavelength magnetothermal drip modes can still proceed at a rate similar to $\omega_{\rm ti}$, even if the magnetic field is purely vertical, as originally shown by \citet{Loewenstein_1990} and \citet{Balbus_1991}. That happens because vertical magnetic fields inhibit horizontal motion of the plasma immediately beneath a low-entropy perturbation that is descending. Compression of the gas beneath a descending low-entropy perturbation therefore provides additional pressure support for it, resulting in an outcome similar to the case of horizontal magnetic fields: Once again, a low-entropy perturbation cools faster than it can descend and steadily gains contrast, relative to its surroundings.  

\section{Magnetohydrodynamic Precipitation}
\label{sec:MHDPrecip}

We are now prepared to explain what the term \textit{magnetohydrodyamic precipitation} is supposed to mean: It describes nonlinear density and entropy contrasts in a stratified galactic atmosphere that have arisen through magnetothermal drip modes. The linearized analysis of section \ref{sec:MTD1} provides some clues that can help distinguish magnetohydrodynamic precipitation from density perturbations that have other origins:
\begin{itemize}

    \item \textbf{Vertical magnetized filaments.} Thermal instability can produce nonlinear density contrasts in a stratified medium, even if $t_{\rm cool} / t_{\rm ff} \sim 10$, as long as magnetic tension can resist the buoyancy effects that would otherwise interfere with thermal instability. If the initial field is primarily horizontal, then magnetothermal drip generates vertical field in the vicinity of the condensates. And if the initial field is primarily vertical, then horizontal displacements strengthen the horizontal field as condensation proceeds and the condensates descend.

    \item \textbf{Characteristic wavenumber.} According to the linearized analysis of magnetothermal drip, the longest wavelength mode that can grow on a timescale similar to $t_{\rm cool}$ has a wavenumber $k_0 \sim (\beta_0 /\lambda_K \lambda_P)^{1/2}$.

    \item \textbf{Characteristic filament length.} Extending the linearized analysis of magnetothermal drip into the nonlinear regime implies that the vertical displacement of a mode with nonlinear density contrast should be similar to the atmosphere's pressure scale height.

    \item \textbf{Slow descent speeds.} Extending the linearized analysis also gives a descent speed $\sim \lambda_P / t_{\rm cool}$ for a density perturbation that is becoming nonlinear, and the predicted speed is smaller than the Keplerian speed of the atmosphere's potential well by a factor $\sim t_{\rm ff} / t_{\rm cool}$. As the perturbation's density contrast continues to grow, its downward acceleration remains sub-Keplerian because of vertical magnetic tension.

    \item \textbf{Amplified magnetic fields.} The instability itself amplifies the ambient magnetic field so that the magnetic pressure becomes similar to thermal pressure (i.e., $\beta \sim 1$) in the vicinity of a nonlinear condensate. Further cooling \textit{within} a condensate can further reduce its thermal support, so that resistance to compression of the densest gas becomes primarily magnetic (i.e., $\beta < 1$).

\end{itemize}

\begin{figure}[t]
    \centering
    \includegraphics[width=0.99\linewidth]{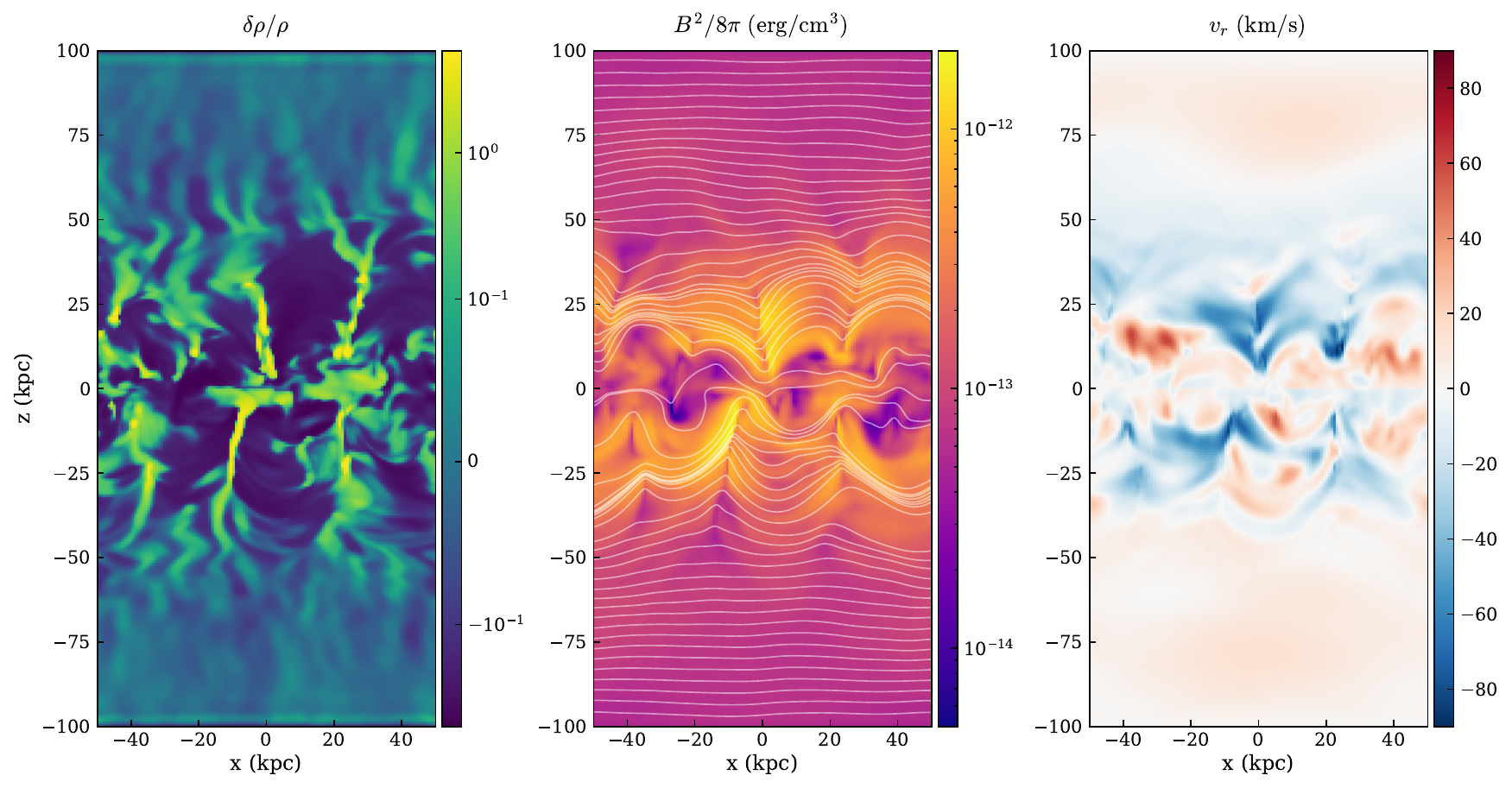}
    \caption{Nonlinear magnetothermal drip filaments in an idealized galactic atmosphere simulated by \citet{Wibking_2025MNRAS.544.2577W}. The left panel shows the density perturbations $(\delta \rho / \rho)$ at the moment of maximum cold-gas accumulation in a 1~kpc thick slice of their simulation D (background atmosphere: $t_\mathrm{cool} / t_\mathrm{ff} \approx 5$, initially horizontal field with $\beta \approx 10^2$). The central panel shows the magnetic energy density $(B^2 / 8 \pi)$ at the same moment, along with examples of individual field lines. The right panel shows velocity ($v_r$) in the direction opposite to $\mathbf{g}$. Yellow regions in the left panel are high-density filaments that grew via thermal instability from small density perturbations (initially with $\delta \rho / \rho \sim 10^{-2})$. Those same regions in the central panel are where magnetothermal drip has amplified the vertical field while producing characteristic kinks in the field lines where the filaments are located. 
    }
    \label{fig:Wibking_MTD}
\end{figure}

Verifying that the linearized analysis accurately predicts what happens in the nonlinear regime requires numerical simulations of magnetothermal drip, and the simulations by \citep{Ji_2018} and \citet{Wibking_2025MNRAS.544.2577W} do, in fact, exhibit examples of the predicted characteristics. Figure \ref{fig:Wibking_MTD} shows how magnetothermal drip manifests in simulation D from \citet{Wibking_2025MNRAS.544.2577W}. That simulation of an isothermal exponential atmosphere with $t_\mathrm{cool} / t_\mathrm{ff} \approx 5$ and a 50~kpc scale height initially has a uniform horizontal magnetic field with $\beta \approx 100$. 

The figure shows a narrow slice, parallel to the initial magnetic field, of the simulation environment at the moment when the mass of condensed gas reaches its maximum value. In the left panel showing density perturbations, the yellow regions are nearly vertical filaments several times denser than the mean density at similar altitudes. In the central panel showing magnetic energy density, the field lines within 25~kpc of the midplane have morphologies resembling the schematic illustration in Figure \ref{fig:MHD_Drip_nonlinear}: The field has been amplified by an order of magnitude near the filaments, becoming nearly vertical in some places, and sharp kinks along the spines of the densest filaments point toward the bottom of the potential well. Also, the lengths of the filaments are similar to the atmosphere's scale height. 

In an isothermal background atmosphere with $\beta \approx 100$ and $\lambda_P = 50$~kpc, the characteristic wavenumber for magnetothermal drip is $k_0 \sim \beta_0^{1/2} / \lambda_P \sim 0.2 \, \mathrm{kpc}^{-1}$. The expected separation between filaments is then $2 \pi / k_0 \sim 30 \, \mathrm{kpc}$, and the filaments in Figure \ref{fig:Wibking_MTD} are indeed separated by roughly that distance. Furthermore, both \citet{Ji_2018} and \citet{Wibking_2025MNRAS.544.2577W} show that reducing $\beta_0$ reduces the characteristic wavenumber of thermal instability and increases the spacing between filaments. 

At the moment shown in Figure \ref{fig:Wibking_MTD}, the densest filamentary gas is descending at speeds up to $\sim 80 \, \mathrm{km \, s^{-1}}$. Such speeds are broadly consistent with our extension of linearized magnetothermal drip, which predicts that the filaments should be descending at speeds $\sim \lambda_P / t_\mathrm{cool}$ when their density contrasts become nonlinear. In this particular simulation, the atmosphere's scale height is 50~kpc and the initial cooling time at 25~kpc is 0.6~Gyr, and so the expected speeds of newly forming filaments descending from that altitude are $\sim 85 \, \mathrm{km \, s^{-1}}$. 

\begin{figure}[t]
    \centering
    \includegraphics[width=0.88\linewidth]{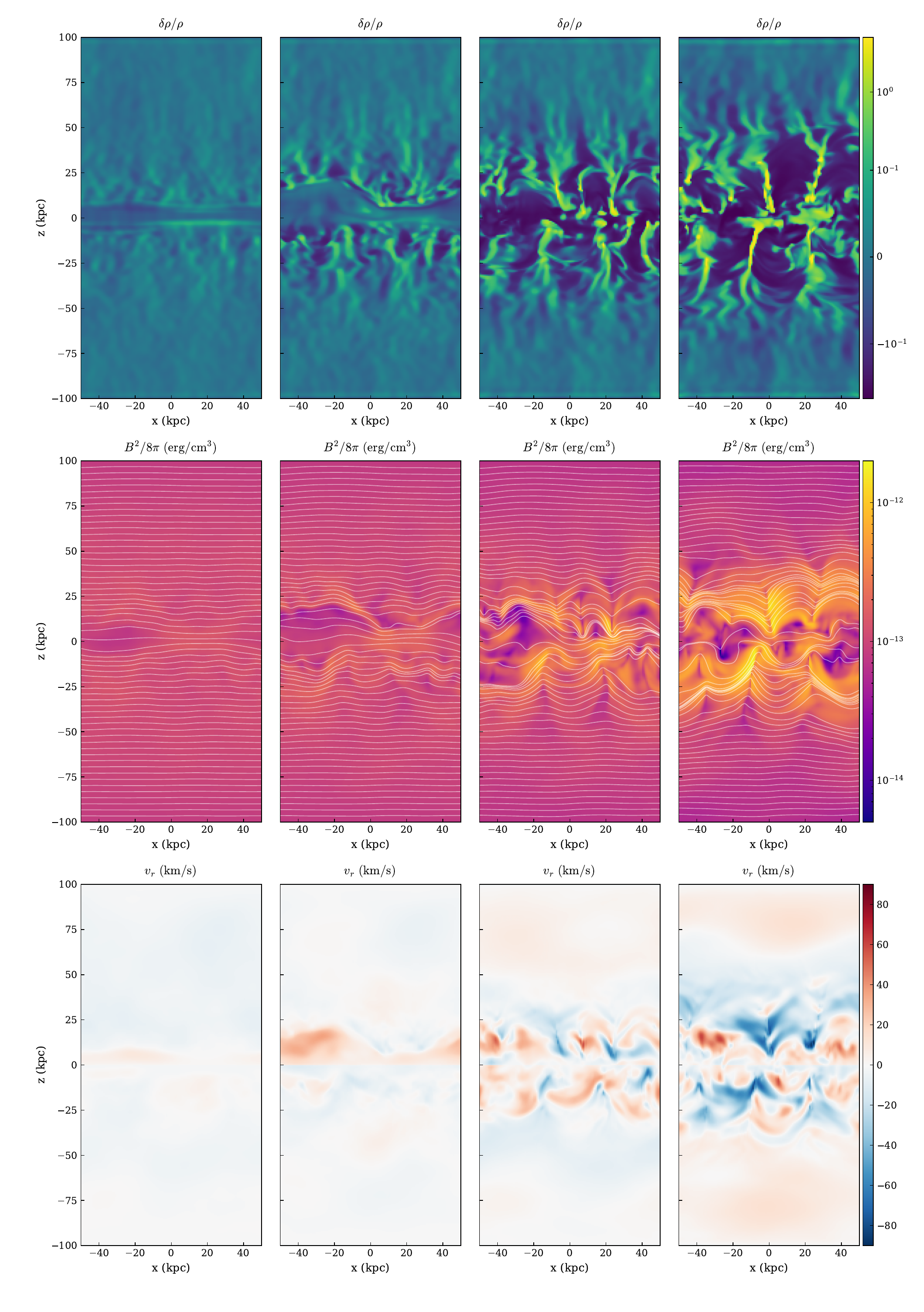}
    \caption{Filament formation in an idealized galactic atmosphere simulated by \citet{Wibking_2025MNRAS.544.2577W}. The panels in the right-hand column are identical to the panels in Figure \ref{fig:Wibking_MTD}, and each row of panels shows how the corresponding quantity evolves.
    }
    \label{fig:FilamentFormation}
\end{figure}

Figure \ref{fig:FilamentFormation} presents a gallery of panels including three previous moments in the simulation that illustrate how the filaments in Figure \ref{fig:Wibking_MTD} develop. Panels in the left column show an early moment, while all perturbations remain of linear order. The magnetic field perturbations are becoming nonlinear in the second column of panels, but the density perturbations are still relatively small. Dense filaments have started to form in the third column, and they are descending at several tens of km~s$^{-1}$. In the fourth column, corresponding to Figure \ref{fig:Wibking_MTD}, the filament system is fully developed up to altitudes of $\sim 30 \, \mathrm{kpc}$.

\begin{figure}[t]
    \centering
    \includegraphics[width=0.99\linewidth]{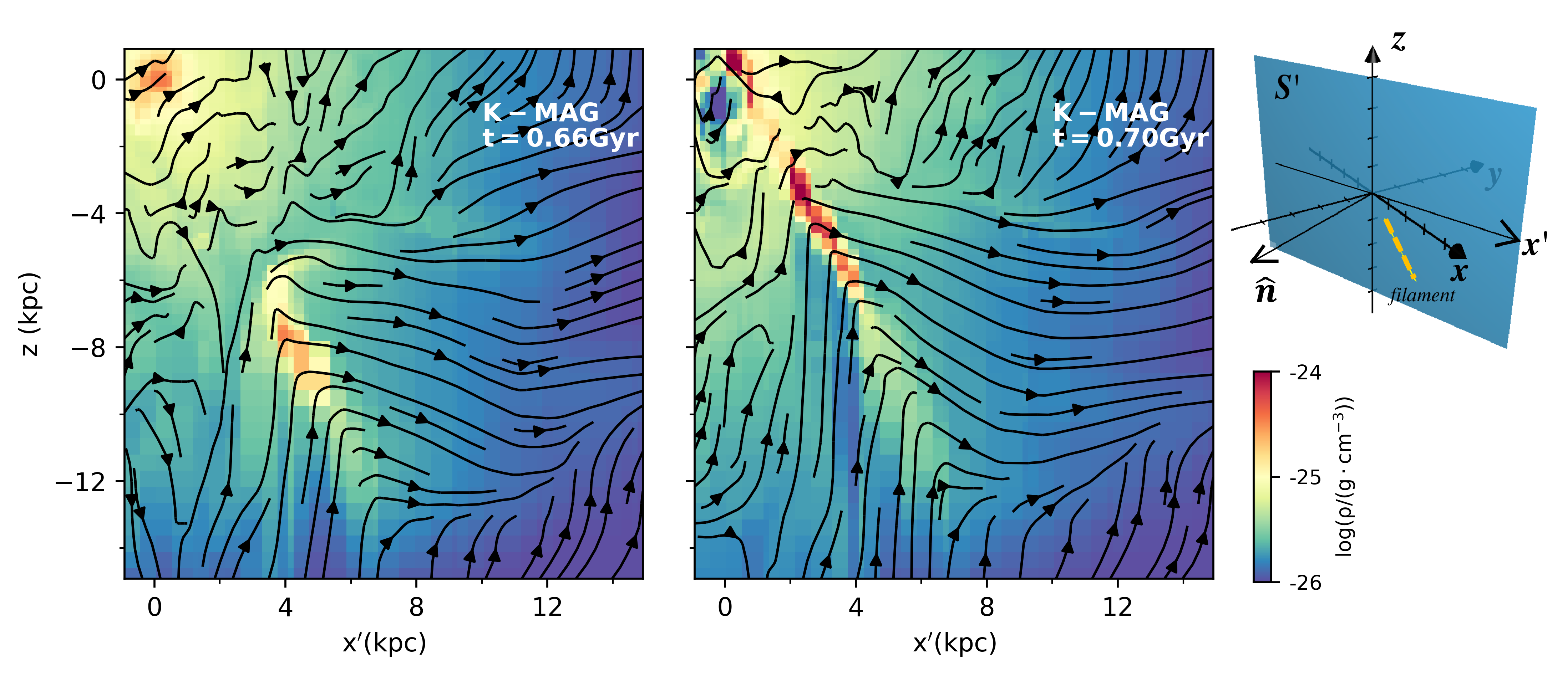}
    \caption{A filament resembling nonlinear magnetothermal drip in a simulation of self-regulating AGN feedback by \citet{WangRuszkowskiYang_2020MNRAS.493.4065W}. The left panel shows gas density and magnetic field morphology at an early stage of filament development. A sharp kink in the field lines, pointing toward the bottom of the potential well, coincides with the densest gas. The right panel shows the same filament at a more mature stage of development, when the kink has become more pronounced and the filament has become more elongated. The $S^\prime$ plane in the 3-D coordinate system at the far right is the filament's plane in simulation coordinates, and an orange line in that plane schematically represents the filament.
    }
    \label{fig:WangFilament}
\end{figure}

There also appear to be examples of magnetothermal drip in simulations that were not explicitly designed to study magnetohydrodynamic precipitation. Figure \ref{fig:WangFilament} shows the magnetic field geometry in the vicinity of a cold, dense filament that has formed in a simulation by \citet{WangRuszkowskiYang_2020MNRAS.493.4065W}. The filament is nearly vertical, meaning that it is aligned with the gravitational potential gradient, and its densest gas is collecting in a magnetic-field kink that points toward the potential well's center, enabling magnetic tension to resist gravitational acceleration. And indeed, during the 40~Myr interval separating the panels the dense gas has descended by $\sim 3$~kpc, implying a sub-Keplerian descent speed $\sim 100 \, \mathrm{km \, s^{-1}}$. Furthermore, magnetic field amplification in the simulation volume is highly correlated with cold, dense gas. Initially, the background plasma has $\beta \sim 10^2$, but where cold gas precipitates, the ratio of thermal to magnetic pressure locally reaches $\beta \sim 1$.

\begin{figure}[t]
    \centering
    \includegraphics[width=0.99\linewidth]{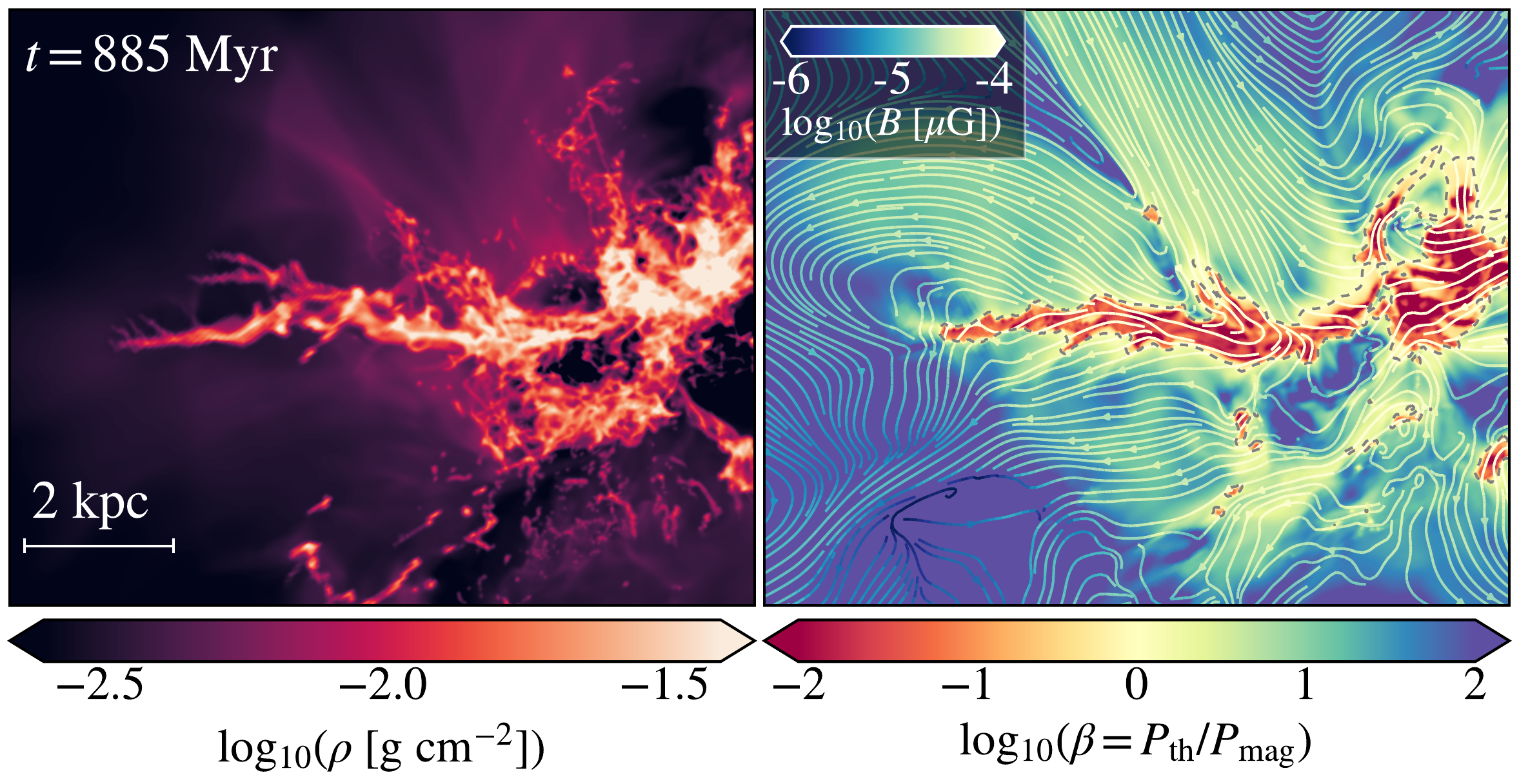}
    \caption{A filament resembling nonlinear magnetothermal drip in a high-resolution cool-core cluster simulation by \citet{Fournier_2024A&A...691A.239F}. The left panel shows gas density in and around their filament \#3, an example of \textit{in situ} condensation that was not obviously stimulated by uplift of low-entropy gas during a feedback outburst. Lines in the right panel show the magnetic field structure in the vicinity of filament \#3; background color-coding shows the local value of $\beta$. The field's morphology resembles the morphology of magnetothermal drip in more idealized simulations, and development of the filament has clearly boosted the local field strength. The background field is weak ($\beta \sim 10^2$), but magnetic pressure close to the filament is comparable to thermal pressure ($\beta \sim 1$), and magnetic pressure dominates thermal pressure ($\beta \ll 1$) in the densest filamentary gas.
    }
    \label{fig:FournierFilament3}
\end{figure}

Figure \ref{fig:FournierFilament3} shows another example of apparent magnetothermal drip in a simulation by \citet{Fournier_2024A&A...691A.239F}. It resides in a simulation environment of considerable complexity. Sporadic eruptions of energy from the black-hole feedback mechanism at the center strongly interact with the filament system that develops. In some cases those eruptions stimulate filament formation by lifting low-entropy gas with a short cooling time to greater altitudes, where it can develop strong density contrasts before descending. However, the filament in Figure \ref{fig:FournierFilament3} does not result from uplift, and its magnetic-field morphology is consistent with magnetothermal drip. Like the filament in Figure \ref{fig:WangFilament}, dense gas is accumulating near a sharp kink in the magnetic field lines, and the magnetic field strength in its vicinity has become amplified by more than an order of magnitude. Also, the majority of the filament's gas is descending at sub-Keplerian speeds ($\sim 100 \, \mathrm{km \, s^{-1}}$).

\begin{figure}[t]
    \centering
    \includegraphics[width=0.99\linewidth]{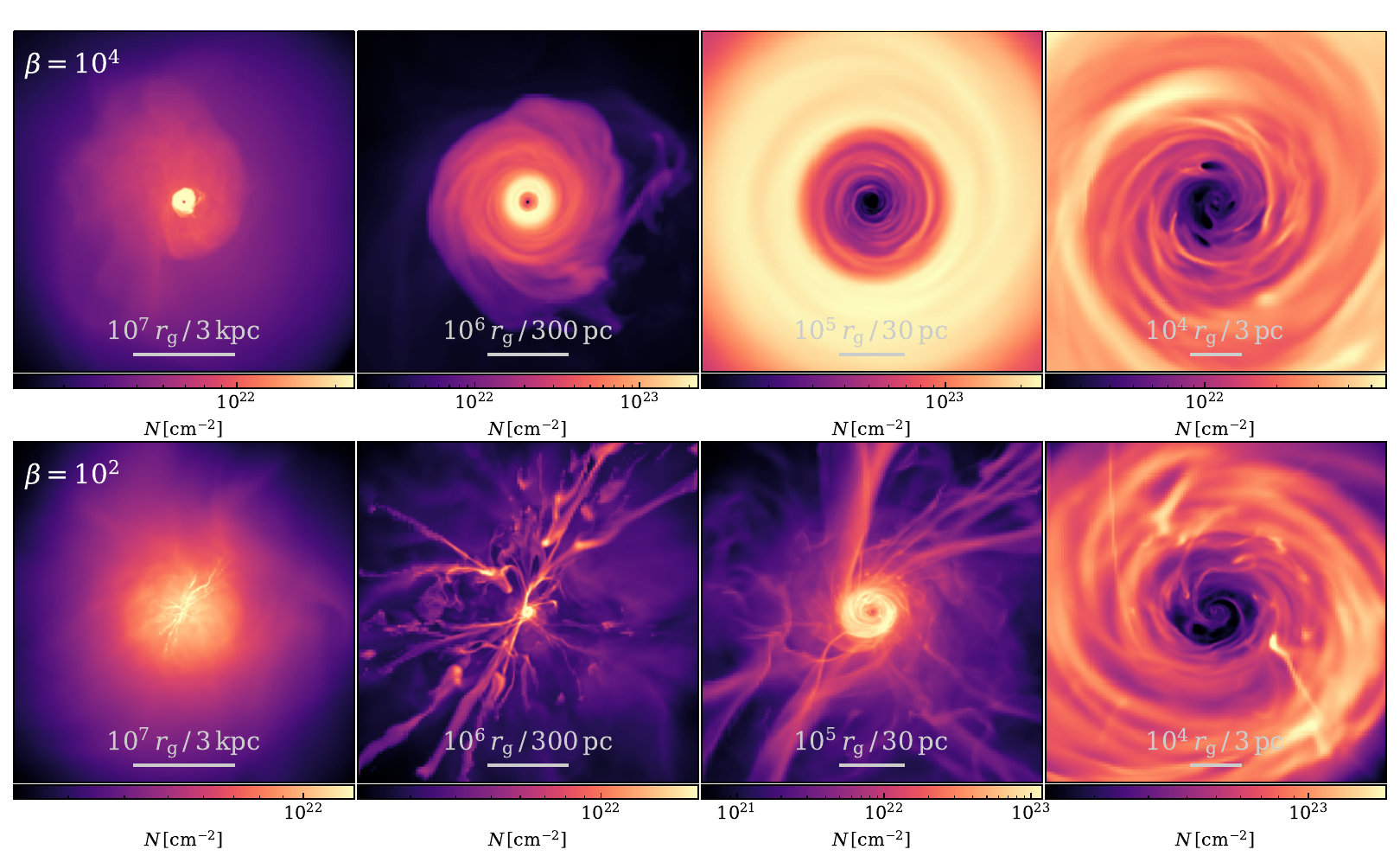}
    \caption{Dependence of filament formation and accretion rate on the background value of $\beta$ in suites of nested magnetohydrodynamic simulations by \citet{Guo_2024ApJ...973..141G} of an idealized M87-like environment. The bottom row of panels shows the suite with an initially weak magnetic field ($\beta = 10^2$). The top row shows the suite with magnetic fields that are initially two orders of magnitude weaker ($\beta = 10^4$). Vertical filaments form on kpc scales in the $\beta = 10^2$ simulation but not in the $\beta = 10^4$ simulation. Also, accretion of dense gas in the $\beta = 10^2$ simulation forms a $\sim 30 \, \mathrm{pc}$ accretion disk that is an order of magnitude smaller, with an accretion rate an order of magnitude greater, than in the $\beta = 10^4$ simulation.
    }
    \label{fig:Guo_fig_proj_beta}
\end{figure}

Figure \ref{fig:Guo_fig_proj_beta} presents a third example. It comes from two simulation suites by \citet{Guo_2024ApJ...973..141G} designed to track cooling-driven accretion onto a supermassive black hole in an M87-like environment. The top row of panels shows how accretion develops in a medium with a magnetic field that is initially randomly oriented and very weak $(\beta = 10^4)$. There are no obvious long radial filaments on kpc scales, and condensing gas settles into a $\sim 300$~pc disk. The bottom row shows a simulation suite with nearly identical initial conditions, except for the magnetic field strength, which is two orders of magnitude greater $(\beta = 10^2)$ but still weak relative to thermal pressure. In that case, long filaments aligned with the gravitational potential gradient form on kpc scales, flow nearly radially inward, and accrete onto a much smaller $\sim 30$~pc disk. The magnetic fields in the cold gas have been amplified by more than an order of magnitude, and the filamentary gas on kpc scales is descending at sub-Keplerian speeds. Also, the accretion rate in the $\beta = 10^2$ simulation is an order of magnitude greater than in the $\beta = 10^4$ simulation, at least in part because of magnetic torques, which we will discuss next.

\section{Magnetic Torques}
\label{sec:Torques}

Magnetothermal drip in a spherically symmetric potential well develops torques that help condensed gas sink to the center of the potential well. Without magnetic fields, the orbital speed of a dense gas cloud with some angular momentum increases as the cloud drifts inward, ultimately causing it to orbit at a radius determined by its specific angular momentum. However, the magnetic field of a cloud resulting from magnetothermal drip will act like a trailing tether opposing both vertical and horizontal motions. The horizontal component of magnetic force therefore removes angular momentum from a descending filament, which ends up falling deeper into the potential well before settling into an orbiting disk.

The magnetohydrodynamic simulations by \citet{WangRuszkowskiYang_2020MNRAS.493.4065W} compellingly illustrate the phenomenon. Prior simulations by \citet{WangLiRuszkowski_2019MNRAS.482.3576W} had explored self-regulating AGN feedback fueled by precipitating gas clouds in massive elliptical galaxies with \textit{unmagnetized} atmospheres. Many features of those simulated galaxies resembled observations of real galaxies, except for the massive kpc-scale disks of cold gas that formed at their centers. And when \citet{WangRuszkowskiYang_2020MNRAS.493.4065W} added magnetic fields to the previous simulations, which were otherwise identical, the anomalous kpc-scale disks vanished. 

To analyze the reason, they evaluated a ``damping time" on which magnetic tension drains momentum from the cold, dense gas clouds that form via precipitation in their simulations. They found a typical damping time of $\sim 10$~Myr, shorter than the freefall time at 1--10~kpc, where most of the clouds were forming. Under those conditions, magnetic tension reduces a cloud's angular momentum as it falls inward, resulting in nearly radial infall, as shown in Figure \ref{fig:WangFilament}.

A simple toy model will help to clarify how magnetothermal drip leads to nearly radial infall. Consider a low-entropy drip perturbation that achieves nonlinear density contrast with a vertical speed $v_\parallel < 0$ and a tangential speed $v_\perp$ initially smaller than $| v_\parallel|$. It is moving at an angle $\theta \sim v_\perp / v_\parallel$ with respect to gravity, and the magnetic field lines it has amplified trail behind it, applying an acceleration comparable to $|\mathbf{g}|$ in the trailing direction. They consequently apply an approximate torque per unit mass $(v_\perp / v_\parallel) g r$ that opposes the falling gas clump's specific angular momentum $v_\perp r$, so that
\begin{equation}
    \frac {d} {dt} v_\perp r 
        \: \approx \: - \frac {v_\perp} {| v_\parallel| } gr
        \; \; .
\end{equation}
Recognizing that $g = v_\mathrm{c}^2/r$ and $d r / dt = v_\parallel$, we can rewrite this expression as
\begin{equation}
    \frac {1} {v_\perp} \frac {d v_\perp} {dt}
        \: \approx \: \frac {|v_\parallel|} {r} 
            \left( 1 - \frac {v_\mathrm{c}^2} {v_\parallel^2} \right)
            \; \; .
\end{equation}
The tangential speed of a developing magnetothermal drip filament therefore \textit{decreases} as it descends until its descent speed approaches the potential well's circular velocity.

Such magnetic torques presumably explain why the \citet{Guo_2024ApJ...973..141G} simulation with $\beta = 10^2$ has a much smaller disk than the one in the simulation with $\beta = 10^4$ (see Figure \ref{fig:Guo_fig_proj_beta}). Filaments with the characteristics of magnetothermal drip form in the simulation with $\beta = 10^2$ on kpc scales with modest amounts of angular momentum. At $\sim 3 \, \mathrm{kpc}$ from the center, the cold, filamentary gas has specific angular momentum $\sim 200 \, \mathrm{kpc \, km \, s^{-1}}$. It would therefore settle into orbits with radii $\sim 0.5 \, \mathrm{kpc}$ in a potential well with $v_\mathrm{c} \sim 400 \, \mathrm{km \, s^{-1}}$, similar to the disk radius in the $\beta = 10^4$ simulation, if there were no torques. However, the specific angular momentum of the infalling filaments in the $\beta = 10^2$ simulation declines to $\sim 30 \, \mathrm{kpc \, km \, s^{-1}}$ as the cold gas sinks to $\sim 200$~pc. At that radius its infall speed approaches $v_\mathrm{c}$, and thereafter the angular momentum of the infalling cold gas remains nearly constant until it flows onto the $\sim 30$~pc disk.

\section{Thermal Conduction}
\label{sec:Conduction}

The preceding calculations have ignored thermal conduction, which can compete with radiative cooling on sufficiently short length scales. To assess its effects, we will define a temperature-dependent conduction coefficient $\kappa(T)$ so that the conductive heat flux is $- \kappa \nabla T$. The local conductive heating rate per unit volume is then $-\nabla \cdot ( \kappa \nabla T )$, and it has a magnitude $k^2 \kappa T$ when $T$ is sinusoidal with wavenumber $k$. Conduction is therefore comparable to radiative cooling for $k^2 \sim \mathcal{L} \rho / \kappa T$, and it redistributes heat more rapidly than radiative cooling can shed heat for larger values of $k$.

Switching to notation that might be more familiar, we can express radiative cooling in terms of a temperature dependent cooling function $\Lambda(T)$, defined so that $n_e^2 \Lambda = \mathcal{L} \rho$. The critical length scale on which conduction and radiative cooling are comparable, called the \textit{Field length} to honor George Field's pioneering work on thermal instability \citep{Field_1965}, is then
\begin{equation}
    \lambda_{\rm F} 
        \equiv \sqrt{ \frac {\kappa T}
                        {n_e^2 \Lambda}} 
        \; \; .
\end{equation}
Conduction is negligible for $k \ll \lambda_{\rm F}^{-1}$, but it suppresses magnetothermal drip (and other forms of thermal instability) for $k  \gtrsim \lambda_{\rm F}^{-1}$.

In highly ionized astrophysical plasmas, the standard conduction coefficient is the Spitzer one:
\begin{equation}
    \kappa (T)  
        \approx ( 5 \times 10^{-7} \, 
            {\rm erg \, s^{-1} \, cm^{-1} \, K^{-7/2}})
            \, f_{\rm c} \, T^{5/2}
            \; \; .
\end{equation}
Here, an arbitrary suppression factor ($f_{\rm c}$) has been included to account for plasma effects capable of reducing conductive heat transport below the classic Spitzer value. For this conduction coefficient, the Field length becomes
\begin{equation}
 \begin{split}
        \approx 3 \, {\rm kpc} \: &
            \left( \frac {K_e} 
                {10 \, {\rm keV \, cm^2}} \right)^{3/2}
                \left( \frac {T} 
                {10^7 \, {\rm K}} \right)^{1/4}
            \left( \frac {\Lambda} 
                {3 \times 10^{-23} \, {\rm erg \, cm^3 \, s^{-1}} } \right)^{-1/2} 
                            f_{\rm c}^{1/2} 
 \end{split}            .
\end{equation}
in which $K_e \equiv kT n_e^{-2/3}$ is a measure of specific entropy commonly used by X-ray astronomers. They find $K_e \sim 10 \, {\rm keV \, cm^2}$ in galaxy cluster cores threaded with much denser filamentary gas \citep[e.g.,][]{Donahue_2006ApJ...643..730D}, and the $T^{1/4}$ and $\Lambda^{-1/2}$ factors nearly offset each other for temperatures exceeding $10^7$~K. Magnetothermal drip is therefore a plausible origin for the filaments observed in those clusters, as long as
\begin{equation}
    \beta_0 
        \lesssim \frac {2 \gamma \lambda_K \lambda_P} 
                        {\lambda_{\rm F}^2}
            \; \; .
\end{equation}
This constraint becomes $\beta_0 \lesssim 56 \, f_{\rm c}^{-1}$ if we assume $K_e \propto r^{2/3}$, $P \propto r^{-1}$, and $K_e \approx 10 \, {\rm keV \, cm^2}$ near $r \sim 10$~kpc.

Keep in mind, however, that magnetic fields strongly suppress the conductive heat flux perpendicular to the field lines, and so steep temperature gradients can persist over length scales much smaller than $\lambda_F$ in two out of three directions. Multiphase filaments resulting from magnetothermal drip can therefore be quite narrow in comparison to their lengths along the field lines, which need to exceed the Field length $\lambda_{\rm F}$ while the filaments are developing.

\section{Summary}
\label{sec:Summary}

This tutorial has explained how weak magnetic fields assist growth of thermal instability in stratified galactic atmospheres. In unmagnetized atmospheres with $t_\mathrm{cool} \gg t_\mathrm{ff}$, the thermally unstable modes are internal gravity waves that saturate before thermal instability produces nonlinear density contrasts and precipitation. However, weak magnetic fields enable a distinctly different thermally unstable mode to grow \citep{Loewenstein_1990, Balbus_1991}. This tutorial calls that mode \textit{magnetothermal drip}. It grows without oscillating because magnetic tension limits the speeds at which overdense regions descend, and thereby enables thermal instability to produce large density contrasts, resulting in \textit{magnetohydrodynamic precipitation}. Numerical simulations show that this unstable mode characteristically produces nearly vertical filaments that descend at sub-Keplerian speeds and strongly amplify the magnetic fields in the vicinity of the densest gas.
\\

The authors acknowledge support from the U.S. National Science Foundation (NSF) through grant AAG-2106575.

\appendix

\section{Saturation of Thermally Unstable Gravity Waves}
\label{app:gMode_Saturation}

In numerical simulations of stratified galactic atmospheres without magnetic fields, thermally unstable $g$-modes with $\omega_{\rm ti} \ll \omega_{\rm BV}$ saturate and their entropy perturbation amplitudes stop growing when the entropy contrasts reach $\delta_K \sim \omega_{\rm ti}/\omega_{\rm BV}$. The saturation amplitude is more commonly expressed as $\delta_\rho \sim A ( t_{\rm cool} / t_{\rm ff} )^{-1}$, where $A$ is a normalization factor of order unity depending on the background atmosphere's characteristics \citep[e.g.,][]{McCourt_2012,Wibking_2025}. Nonlinear effects in stratified atmospheres with $t_{\rm cool}/t_{\rm ff} \gg 1$ therefore usually halt growth of thermal instability before the perturbations achieve nonlinear density amplitudes \citep[but see][]{Choudhury_2019MNRAS.488.3195C}. 

Saturation happens when nonlinear coupling of $g$-modes becomes comparable to thermal pumping of mode growth \citep[e.g.,][]{McCourt_2012,Voit_2017,Voit_2021ApJ...913..154V}. A gravity-wave perturbation oscillating with frequency $\sim \omega_{\rm BV}$ moves vertically at a speed $\sim  \omega_{\rm BV} | \xi_\parallel |$. Nonlinear mode coupling therefore drains energy from it at a rate $\sim  \omega_{\rm BV} | k \xi_\parallel |$, where $k$ is the mode's wavenumber. That rate is similar to $\omega_{\rm ti}$ for $| \xi_\parallel | \sim \omega_{\rm ti} / k \omega_{\rm BV}$, at which point the perturbation's entropy contrast has grown to be
\begin{equation}
    | \delta_K | \sim \frac {|\xi_\parallel|} {\lambda_K}
        \sim ( k \lambda_K )^{-1} \frac {\omega_{\rm ti}} {\omega_{\rm BV}}
        \; \; .
\end{equation}
The longest wavelength modes, with $k \lambda_K \sim 1$, consequently produce the greatest entropy contrasts. They are responsible for the largest density perturbations seen in simulations with $\omega_{\rm ti} \ll \omega_{\rm BV}$ and have $\delta_\rho \sim ( t_{\rm cool} / t_{\rm ff} )^{-1} \ll 1$.

\section{Precipitation and Freefall}
\label{app:Freefall}

Nonlinear saturation of thermal instability, as outlined in Appendix \ref{app:gMode_Saturation}, depends on stratification and buoyancy. The rate at which buoyancy acts depends on the effective gravitational acceleration $\mathbf{\tilde{g}}$. If the background medium is static, then $\mathbf{\tilde{g}}$ equals the actual gravitational acceleration $\mathbf{g}$.  In that case, $\tilde{\omega}_{\rm BV}$ is equal to $\omega_{\rm BV}$. But if the background medium is accelerating, the saturation amplitude becomes
\begin{equation}
    | \delta_K | 
        \sim ( k \lambda_K )^{-1} \frac {\omega_{\rm ti}} {\tilde{\omega}_{\rm BV}}
        \; \; .
\end{equation}
Importantly, both $\mathbf{\tilde{g}}$ and $\tilde{\omega}_{\rm BV}$ vanish in a freely falling background medium, allowing thermally unstable entropy contrasts to progress to nonlinear amplitudes without saturating, regardless of their physical size.

Assessments of the critical value of $t_{\rm cool}/t_{\rm ff}$ for development of multiphase gas therefore need to consider whether the background medium could be accelerating in the direction of $\mathbf{g}$. Observations of pressure gradients in the plane of the sky provide useful information for that purpose, because $\rho \mathbf{\tilde{g}} = \nabla P$. An atmosphere observed to have $| \nabla P | \ll \rho \mathbf{g}$ may well be close to freefall and is then susceptible to developing multiphase structure via thermal instability. However, nearly homogeneous gas with $t_{\rm cool} / t_{\rm ff} \gg 1$ would need to remain in freefall for multiple orbital periods before achieving nonlinear entropy contrasts.

\section{Thermal Stability of Acoustic Modes}
\label{app:AcousticStability}

The tutorial's main text focuses on three thermally unstable normal modes of a stratified atmosphere with $t_{\rm cool} > t_{\rm ff}$: one magnetothermal drip mode and two magneto-gravity modes. All three modes are thermally unstable in the linear regime. However, the dominant modes depend on wavelength. Long wavelength $(k < k_0)$ magneto-gravity modes grow faster than long-wavelength drip modes, and short-wavelength drip modes ($k > k_0$) grow faster than magneto-gravity modes with similarly short wavelengths.

A magnetohydrodynamic atmosphere's remaining normal modes are magneto-acoustic modes, analogous to fast and slow magnetosonic waves, but with a gravitational correction that becomes increasingly important for long-wavelength modes with frequencies approaching $\tilde{\omega}_{\rm BV}$. We implicitly set aside those acoustic modes in sections \ref{sec:PressurePerts} and \ref{sec:MTD2} when we assumed that $\omega^2 \ll k_\perp^2 c_{\rm s}^2$. This appendix analyzes these acoustic modes and takes a modest step toward completeness by assessing their thermal stability in the limit of negligible magnetic field ($\beta_0 \gg 1$). It shows that thermally unstable acoustic modes, if present, grow slowly compared to thermally unstable entropy perturbations.

To start, we will go back to equation (\ref{eq:omega2}) and use equation (\ref{eq:deltaP_gamma}) to substitute for $\delta_P/\gamma$, giving us a differential equation with the form
\begin{equation}
    \frac {\partial^2 \xi_\parallel} {\partial \tilde{r}^2}
        + f_1 (\tilde{r}) 
            \frac {\partial \xi_\parallel} {\partial \tilde{r}}
        + f_2 (\tilde{r})  \xi_\parallel
        = 0
        \label{eq:xiparallel_diffeq}
        \; \; .
\end{equation}
The complexities that vertical stratification and thermal instability add to acoustic modes have been bundled into two functions of $\tilde{r}$:
\begin{equation}
    f_1 (\tilde{r}) 
        = - \frac {1} {\lambda_P} 
            + \frac {1} {\lambda_{\rm s}} 
            + \left( \frac {i \omega_{\rm ti}} {\omega - i \omega_{\rm ti}} \right)
                \frac {1} {\gamma \lambda_K}
                \; \; ,
\end{equation}
\begin{equation}
    f_2 (\tilde{r}) 
        = \left( \frac {\omega^2} {c_{\rm s}^2} - k_\perp^2 \right) 
               \left[ 1 - \left( \frac {\omega} {\omega - i \omega_{\rm ti}} \right) 
                    \frac {\tilde{\omega}_{\rm BV}^2} {\omega^2} \right]
                + \left( \frac {\gamma - 1} {\gamma \lambda_P} 
                    - \frac {1} {\lambda_{\rm s}} \right) \frac {1} {\lambda_\Delta}
                    - \frac {d} {d \tilde{r}} \frac {1} {\lambda_\Delta} 
    \; \; .
\end{equation}
Notice that two new inverse length scales appear in these equations for $f_1$ and $f_2$:
\begin{equation}
    \frac {1} {\lambda_\Delta} 
        \equiv \frac {1} {\gamma \lambda_P} + \frac {\Delta_K} {\gamma \xi_\parallel} 
        = \frac {\tilde{g}} {c_{\rm s}^2} 
            - \left( \frac {i \omega_{\rm ti}} {\omega - i \omega_{\rm ti}} \right) \frac {1} {\gamma \lambda_K} 
            \; \; ,
\end{equation}
\begin{equation}
    \frac {1} {\lambda_{\rm s}}
        \equiv \left( \frac  {k_\perp^2 c_{\rm s}^2} 
            {\omega^2 - k_\perp^2 c_{\rm s}^2} \right)
        \frac {d} {d \tilde{r}} \ln c_{\rm s}^2
        \; \; .
\end{equation}
The inverse length scale $1 / \lambda_\Delta$ accounts for compression resulting from both the ambient pressure gradient (via $\lambda_P$) and entropy gains or losses (via $\Delta_K$) as a fluid element is vertically displaced. The inverse length scale $1/\lambda_{\rm s}$ accounts for vertical changes in the background sound speed $c_{\rm s}^2$, and it is negligible in a nearly isothermal atmosphere. 

To go further, the differential equation for $\xi_\parallel$ can be reduced as follows \citep[e.g.,][]{Tolstoy_1963}:
\begin{equation}
    \frac {\partial^2 h} {\partial \tilde{r}^2}
        + f(\tilde{r}) h
        = 0
    \; \; \; , \; \; \; 
    h \equiv  \xi_\parallel 
                \exp \left[ \frac {1} {2} 
            \int f_1 (\tilde{r}) \, d \tilde{r} \right]
    \; \; \; , \; \; \; 
    f (\tilde{r}) 
        = f_2 (\tilde{r}) - \frac {1} {4} f_1^2 (\tilde{r})
            - \frac {1} {2} \frac {d f_1} {d \tilde{r}}
            \; \; \; .
\end{equation}
The function $f (\tilde{r})$, if it is positive, can be interpreted as the square of a potentially altitude-dependent vertical wavenumber $k_\parallel = \sqrt{f}$. But if $f$ is negative, then $\sqrt{f}$ is imaginary, and the corresponding normal mode has an exponential dependence on $\tilde{r}$. 

Now we will deploy some approximations in order to gain physical insight. We will consider a nearly adiabatic, isothermal, and exponential atmosphere, in which $\omega_{\rm ti}$, $1 / \lambda_{\rm s}$, and $d \lambda_P / d \tilde{r}$ are all negligible, thereby obtaining
\begin{equation}
    f_1 = - \frac {1} {\lambda_P}
    \; \; \; \; , \; \; \; \; 
    f_2 = \frac {\omega^2} {c_{\rm s}^2} - k_\perp^2 + k_\perp^2 \frac {\tilde{\omega}_{\rm BV}^2} {\omega^2}
\end{equation}
\begin{equation}
    f = \frac {\omega^2} {c_{\rm s}^2} - k_\perp^2 + k_\perp^2 \frac {\tilde{\omega}_{\rm BV}^2} {\omega^2}
            - \frac {1} {4 \lambda_P^2}
    \; \; \; , \; \; \; 
    \xi_\parallel (\tilde{r}) \propto e^{\tilde{r}/2 \lambda_P} e^{i \tilde{r} \sqrt{f}}
            \; \; .
\end{equation}
A normal mode with $f > 0$ then has a total wavenumber $k \equiv \sqrt{f + k_\perp^2}$ that is independent of altitude, and we arrive at a classic result, familiar to people who have studied waves in stratified terrestrial or stellar atmospheres: 
\begin{equation}
    k^2 = \frac {\omega^2} {c_{\rm s}^2}
          \left( \frac{\omega^2 - \omega_{\rm ac}^2} {\omega^2 - \omega_{\rm buoy}^2} \right)
          \label{eq:k_squared}
          \; \; .
\end{equation}
The frequency $\omega_{\rm ac}$ in this expression is known as the atmosphere's \textit{acoustic cutoff}. It separates the atmosphere's normal modes into three categories:
\begin{itemize}
    \item Low-frequency modes with $\omega^2 < \omega_{\rm buoy}^2 < \omega_{\rm ac}^2$ are buoyancy-driven gravity waves ($g$-modes) that are periodic in $\tilde{r}$. 
    \item High-frequency modes with $\omega^2 > \omega_{\rm ac}^2 > \omega_{\rm buoy}^2$ are pressure-driven acoustic waves ($p$-modes) that are periodic in $\tilde{r}$. 
    \item Modes with intermediate frequencies ($\omega_{\rm ac}^2 > \omega^2 > \omega_{\rm buoy}^2$) have $f < 0$ and are called  \textit{evanescent} because of their exponential dependence on altitude. Whether or not they are present depends on boundary conditions at the top and bottom of the atmosphere. 
\end{itemize}
Given the assumptions we have made, this idealized atmosphere's acoustic cutoff is at $\omega_{\rm ac}^2 = c_{\rm s}^2 / 4 \lambda_P^2 = [\gamma^2 / 4 (\gamma - 1)] \, \tilde{\omega}_{\rm BV}^2$. However, $c_{\rm s}$ and $\lambda_P$ generally do not remain constant with altitude, and $\omega_{\rm ac}^2$ then depends on additional atmospheric characteristics.

We would like to know what happens to those modes when they are not quite adiabatic. To address that question, especially for modes with significant compression, we need to bring back in the dependence of radiative cooling on pressure perturbations. We can do that by defining the frequency
\begin{equation}
    \hat{\omega} \equiv \omega - i \omega_{\rm ti} - i \omega_{\rm sw} \frac {\delta_P} {\delta_K}
    \; \; .
\end{equation}
This frequency is essentially the real part of $\omega$, representing the oscillation frequency with thermal instability removed, for modes that oscillate quickly compared to the cooling time ($\omega^2 \gg \omega_{\rm ti}^2 , \omega_{\rm sw}^2$). With this definition for $\hat{\omega}$, retracing our steps to calculate $f_1$, $f_2$, $f$, and $k^2$ leads to
\begin{equation}
    k^2 = \frac {\hat{\omega} \omega} {c_{\rm s}^2}
          \left( \frac{\omega^2 - \hat{\omega}_{\rm ac}^2} 
                {\hat{\omega} \omega - \omega_{\rm buoy}^2} \right)
    \; \; \; , \; \; \;
    \hat{\omega}_{\rm ac}^2 
        = \omega_{\rm ac}^2 \left[ 1 - \frac {2 (\gamma - 1)} {\gamma}          \left( \frac {\omega - \hat{\omega}} {\hat{\omega}} \right) \right]
          \; \; ,
\end{equation}
in which we have retained only the lowest-order terms in $(\omega - \hat{\omega})/\hat{\omega}$. Writing the result as a polynomial in $\omega$ gives us
\begin{equation}
    \hat{\omega} \omega^3 
        - \left( k^2 c_{\rm s}^2 + 
            \hat{\omega}_{\rm ac}^2 \right) \hat{\omega} \omega
        + k_\perp^2 c_{\rm s}^2 \tilde{\omega}_{\rm BV}^2
        = 0
        \; \; .
\end{equation}
For slowly oscillating modes $(k^2 c_{\rm s}^2 \gg \hat{\omega} \omega)$, this polynomial reduces to
\begin{equation}
    \omega^2 
     - i \left( \omega_{\rm ti} 
        + \omega_{\rm sw} \frac {\delta_P} {\delta_K} \right) \omega
        - \left( \frac {k_\perp^2 c_{\rm s}^2 } {k^2 c_{\rm s}^2  
            + \hat{\omega}_{\rm ac}^2} \right) \tilde{\omega}_{\rm BV}^2
            = 0
            \; \; ,
\end{equation}
implying that those modes grow at a rate close to $\omega_{\rm ti} + \omega_{\rm sw} (\delta_P / \delta_K)$. For rapidly oscillating modes $(\omega^2 \gg \omega_{\rm ac}^2)$, we find 
\begin{equation}
    \omega^2 = k^2 c_{\rm s}^2 + \hat{\omega}_{\rm ac}^2    
    \; \; ,
\end{equation}
which has the approximate square roots
\begin{equation}
    \omega 
        \approx \pm \, k c_{\rm s} 
            \left( 1 + \frac {\omega_{\rm ac}^2} {2 k^2 c_{\rm s}^2}\right)
            \pm i \left( \frac {\gamma-1} {\gamma} \right)
                \frac {\omega_{\rm ac}^2} {k^2 c_{\rm s}^2} 
            \left( \omega_{\rm ti} + \omega_{\rm sw} 
                    \frac {\delta_P} {\delta_K} \right)
            \; \; .
        \label{eq:omega_fast_approx}
\end{equation}
Growth or decay of the fast modes is therefore suppressed by the factor $\omega_{\rm ac}^2 / k^2 c_{\rm s}^2$, which is small. 

Both of these results depend on the perturbation amplitude ratio $\delta_P / \delta_K$ of a particular mode, which we can evaluate as if the mode were adiabatic ($\Delta_K = 0$), because we are currently neglecting higher-order terms in $(\omega - \hat{\omega})/\hat{\omega}$. To do that, we combine the adiabatic result $\delta_K = - \xi_\parallel / \lambda_K$ with equation (\ref{eq:deltaP_gamma}), giving
\begin{equation}
    \frac {\delta_P} {\delta_K}
        = \frac {\omega^2} {\omega^2 - k_\perp^2 c_{\rm s}^2}
            \left( \frac {\gamma \lambda_K} {\xi_\parallel}
                \frac {\partial \xi_\parallel} {\partial \tilde{r}}
                    - \frac {\lambda_K} {\lambda_P} \right)
                    \; \; .
\end{equation}
This ratio is small for slow modes with $k_\perp^2 c_{\rm s}^2 \gg \omega^2$, confirming that the contribution of pressure perturbations to thermal instability can be neglected in that limit. For fast acoustic modes with $k^2 > k_\perp^2$, the derivative $\partial \ln \xi_\parallel / \partial \tilde{r}$ is essentially $i$ times the vertical wavenumber, and so that term does not contribute to thermal instability. The real part of $\delta_P / \delta_K$ is then similar to the ratio of scale heights $\lambda_K /  \lambda_P$, which is equal to $1 / ( \gamma - 1)$ in an isothermal atmosphere, and so it does not compensate for the suppression factor $\omega_{\rm ac}^2 / k^2 c_{\rm s}^2$ in equation (\ref{eq:omega_fast_approx}). In other words, even if fast acoustic modes are thermally unstable, they grow more slowly than thermally unstable $g$-modes.

\section{The ``Entropy Mode"}
\label{app:EntropyMode}

The equations of magnetohydrodynamics yield seven normal modes. Six of them receive nearly all of the attention: two Alfv\'en waves, two fast magnetosonic waves, and two slow magnetosonic waves. The seventh normal mode is called an \textit{entropy mode} or \textit{entropy wave} in the magnetohydrodynamics literature \citep[e.g.,][]{Goedbloed_2019mlap.book.....G}. 

Entropy perturbations in a uniformly magnetized plasma that starts out motionless, isobaric, and thermally stable, without any external forces acting on it, will remain motionless. They have the property $\delta_K = - \gamma \delta_\rho$ and can be spectrally decomposed into a superposition of modes with different wavenumbers but zero frequency. These modes are the entropy modes of magnetohydrodynamics, and they are usually neglected when they do nothing else but remain motionless.

However, buoyancy in the presence of gravity puts entropy modes into action. They become internal gravity waves in an unmagnetized atmosphere with an entropy gradient. And in a magnetized atmosphere, internal gravity waves can blend with Alv\`en waves that have vertical displacements to become magneto-gravity waves. One more degree of freedom enables an entropy mode to become a magnetothermal drip mode: The specific entropy of a fluid element needs to be able to change (i.e. $\dot{\Delta}_K \neq 0$). The fluid element's entropy contrast ($\delta_K$), vertical displacement ($\xi_\parallel$), and magnetic support can all then grow in unison on a timescale similar to $\omega_{\rm ti}$.

\section{Relationships to Previous Work}
\label{app:PreviousWork}

To streamline the tutorial, we chose to postpone comparisons with previous work on magnetized thermal instability until this final section of the appendix. Here are the perspectives our analysis provides on its some of its predecessors:

\begin{itemize}

    \item \citet{Loewenstein_1990} initiated the subject with a linearized analysis of thermal instability in a stratified medium threaded by a uniform magnetic field. Taking an Eulerian approach to the equations of magnetohydrodynamics, he derived a seventh-order polynomial dispersion relation. He then considered two special cases, one having $\mathbf{k}$ parallel to $\mathbf{B}_0$ and another with $\mathbf{k}$ perpendicular to $\mathbf{B}_0$. In the low-frequency limit ($\omega \ll k c_{\rm s}$) of waves propagating parallel to the magnetic field he found the following approximate frequency solution in the limit $\omega_{\rm A} \gg \omega_{\rm BV}$:
    \begin{equation}
        \omega \approx i \omega_{\rm ti} 
            \left( \frac {\omega_{\rm A}^2} {\omega_{\rm A}^2 + \omega_{\rm buoy}^2} \right)
            \; \; .
    \end{equation}
    Here, we have expressed that solution in terms of this article's notation and have suppressed terms related to thermal conduction to make it obvious that this is the characteristic frequency of the mode we have been calling magnetothermal drip. Loewenstein also found that a similar frequency, without a real part, did not appear for waves with $\mathbf{k}$ perpendicular to $\mathbf{B}_0$. He correctly reasoned from these findings that thermal instability can grow without oscillating when magnetic tension and buoyancy forces are nearly in balance. However, his article did not demonstrate that the non-oscillating mode (akin to the entropy mode of magnetohydrodynamics) is distinct from the thermally unstable magneto-gravity modes (akin to Alfv\'en waves).

    \item \citet{Balbus_1991} soon followed up with a Lagrangian analysis of magnetothermal instability. It used the Boussinesq approximation to exclude acoustic modes and showed that the remaining modes could grow at a rate similar to $\omega_{\rm ti}$. Balbus emphasized that growth of thermal instability would proceed at the same rate even for a purely radial background field because of how radial fields restrict horizontal motions. But neither he nor Loewenstein considered how nonlinear saturation might hinder growth of thermal instability before nonlinear density contrasts can develop.

    \item \citet{Ji_2018} performed the first numerical simulations explicitly designed to follow the thermally unstable modes analyzed by Lowenstein and Balbus into the nonlinear regime. As discussed in the main text (\S \ref{sec:MTD1}), their simulations showed that nonlinear saturation limits the density contrasts achieved by thermal instability to
    \begin{equation}
        \frac {\delta \rho} {\rho}
            \sim 3 \beta_0^{-1/2}
                \left( \frac {t_{\rm cool}} 
                    {t_{\rm ff}} \right)^{-1}
        \; \; .
    \end{equation}
    They also showed that thermal instability in the presence of both horizontal and vertical magnetic fields saturated with the essentially the same density contrasts. However, \citet{Ji_2018} argued that the most prominent thermally unstable mode arose from equivalence between the cooling time and an Alfv\'en wave crossing time, implying a critical wavenumber $\sim 1 / v_{\rm A} t_{\rm cool}$ instead of the critical wavenumber $k_0 \sim \beta_0^{1/2} ( \lambda_{\rm K} \lambda_{\rm P})^{-1/2}$ at which magnetic stresses nearly balance buoyant acceleration. 

    \item \citet{WangRuszkowskiYang_2020MNRAS.493.4065W} performed spherically symmetric numerical simulations showing that thermal instability in a weakly magnetized atmosphere with initially tangled fields produces vertical filaments with coherent magnetic fields that resist gravitational acceleration. They found strong spatial correlations between cold, dense filaments and nonlinear magnetic field amplification, and also showed that magnetic tension significantly inhibited descent of the dense gas (see Figure \ref{fig:WangFilament}). 

\end{itemize}

This tutorial's analysis therefore confirms several previous findings: Section \ref{sec:MTD1} shows that $k_0$ is a critical wavenumber for magnetothermal drip, in agreement with the reasoning of \citet{Loewenstein_1990}. Section \ref{sec:Saturation} explains how magnetothermal drip produces filaments like those seen in numerical simulations \citep[e.g.,][]{WangRuszkowskiYang_2020MNRAS.493.4065W, Fournier_2024A&A...691A.239F, Wibking_2025MNRAS.544.2577W}. And section \ref{sec:MTD2} explicitly shows how vertical fields can promote thermal instability by restricting horizontal motions, in agreement with both \citet{Loewenstein_1990} and \citet{Balbus_1991}, and as verified by the simulations of \citet{Ji_2018}. 

However, nonlinear saturation of magnetothermal drip arises from an interplay between radiative cooling and magnetic stresses different from what \citet{Ji_2018} imagined. Section \ref{sec:Saturation} argues that a linear model for the vertical displacement of magnetothermal drip remains a good approximation even when $k_\perp \xi_\parallel \gtrsim 1$, meaning that drip modes with $k \gtrsim k_0$ continue to grow at a rate $\sim \omega_{\rm ti}$. However, the magnetic fields induced by growing displacements produce magnetic pressure gradients \textit{transverse} to those displacements. The resulting transverse accelerations are second order in $|\boldsymbol{\xi}|$ and saturate when $k_0$ times the displacement amplitude parallel to $\mathbf{B}_0$ is $\sim 3 \omega_{\rm ti} / \omega_{\rm A}$.

\section{Glossary of Symbols}
\label{app:Glossary}
\vspace*{2em}

    \begin{tabular}{cl}
      \hline 
       \textbf{Symbol} & \textbf{Description} \\
       \hline
         $A$ 
         & Normalization factor for saturated perturbations
         \\
         $\mathbf{a_B}$ 
         & Magnetic acceleration
         \\
         $a_{\mathbf{B},\parallel}$ 
         & Magnetic acceleration $\parallel$ to $\mathbf{\tilde{g}}$
         \\
         $\mathbf{B}$ 
         & Magnetic field
         \\
         $B_\parallel$ 
         & Magnetic field component $\parallel$ to $\mathbf{\tilde{g}}$
         \\
         $B_\perp$ 
         & Magnetic field component $\perp$ to $\mathbf{\tilde{g}}$
         \\
         $\mathbf{B}_0$ 
         & Unperturbed magnetic field
         \\
         $B_0$ 
         & Magnitude of unperturbed magnetic field
         \\
         $B_{0,\parallel}$ 
         & Component of unperturbed magnetic field $\parallel$ to $\mathbf{\tilde{g}}$
         \\
         $B_{0,\perp}$ 
         & Component of unperturbed magnetic field $\perp$ to $\mathbf{\tilde{g}}$
         \\
         $\mathcal{C}$ 
         & Cooling rate per unit mass
         \\
         $c_{\rm s}$ 
         & Sound speed: $\gamma P_0 / \rho_0$
         \\
         $f$ 
         & Generalized wavenumber parallel to $\mathbf{\tilde{g}}$ 
         \\
         $f_1$ 
         & Damping function for vertical displacements 
         \\
         $f_2$ 
         & Wavenumber function for vertical displacements 
         \\
         $f_{\rm c}$ 
         & Suppression factor for thermal conduction (relative to Spitzer) 
         \\
         $\mathbf{g}$ 
         & Gravitational acceleration 
         \\
         $\mathbf{\tilde{g}}$ 
         & Effective gravitational acceleration in comoving frame: $\mathbf{g} - d \mathbf{\tilde{v}} / dt$
         \\
         $\tilde{g}$ 
         & Magnitude of effective gravitational acceleration 
         \\
         $\mathcal{H}$ 
         & Heating rate per unit mass
         \\
         $K$ 
         & Adiabatic constant / Entropy index: $P/\rho^\gamma$
         \\
         $K_0$ 
         & Unperturbed adiabatic constant / Entropy index: $P_0/\rho_0^\gamma$
         \\
         $K_e$ 
         & Version of $K$ defined with respect to electron density: $kTn_e^{-2/3}$
         \\
         $\mathbf{k}$ 
         & Wavevector
         \\
         $k$ 
         & Total wavenumber (also used to represent Boltzmann's constant)
         \\
         $k_\parallel$ 
         & Wavenumber component $\parallel$ to $\mathbf{\tilde{g}}$ 
         \\
         $k_\perp$ 
         & Wavenumber component $\perp$ to $\mathbf{\tilde{g}}$ 
         \\
         $k_0$ 
         & Critical wavenumber for magnetothermal drip modes 
         \\
         $k_B$ 
         & Wavenumber component $\parallel$ to $\mathbf{B}_0$ 
         \\
         $\mathcal{L}$ 
         & Net cooling rate per unit mass: $\mathcal{C - H}$
         \\
         $m_p$ 
         & Proton mass
         \\
         $n_e$ 
         & Electron density
         \\
         $P$ 
         & Thermal pressure
         \\
         $P_0$ 
         & Unperturbed thermal pressure
         \\
         $q$ 
         & Heat energy per particle
         \\
         $\mathbf{r}$ 
         & Location of a fluid element
         \\
         $\mathbf{r}_0$ 
         & Initial location of a fluid element
         \\
         $\tilde{r}$ 
         & altitude coordinate in direction opposite to $\mathbf{\tilde{g}}$
         \\
         $s$ 
         & Specific entropy
         \\
         \hline
    \end{tabular}

    \begin{tabular}{cl}
      \hline 
       \textbf{Symbol} & \textbf{Description} \\
       \hline
         $T$ 
         & Gas temperature
         \\
         $t$ 
         & Time coordinate
         \\
         $t_{\rm cool}$ 
         & Cooling time: $u/\mathcal{C} \rho$
         \\
         $t_{\rm heat}$ 
         & Heating time: $u/\mathcal{H} \rho$
         \\
         $u$ 
         & Thermal energy density
         \\
         $\mathbf{v}$ 
         & Flow velocity
         \\
         $\mathbf{v}_0$ 
         & Unperturbed flow velocity
         \\
         $v_{\rm A}$ 
         & Alfv\'en speed: $B_0^2/4 \pi \rho_0$
         \\
         $v_{\rm c}$ 
         & Circular velocity of potential well: $(gr)^{1/2}$
         \\
         $\beta$ 
         & Ratio of thermal pressure to magnetic pressure: $8 \pi P / B^2$
         \\
         $\beta_0$
         & Unperturbed ratio of thermal pressure to magnetic pressure: $8 \pi P_0 / B_0^2$
         \\
         $\mathbf{\Delta B}$
         & Magnetic field perturbation: $\mathbf{B} - \mathbf{B}_0$
         \\
         $\Delta_K$
         & Lagrangian entropy perturbation: $\ln [ K(\mathbf{r}_0,t) / K_0(\mathbf{r}_0,t)]$
         \\
         $\Delta_P$
         & Lagrangian pressure perturbation: $\ln [ P(\mathbf{r}_0,t) / P_0(\mathbf{r}_0,t)]$
         \\
         $\Delta_\rho$
         & Lagrangian compression: $- \nabla \cdot \boldsymbol{\xi}$
         \\
         $\delta_K$
         & Eulerian entropy perturbation: $\ln [ K(\mathbf{r},t) / K_0(\mathbf{r},t)]$
         \\
         $\delta_P$
         & Eulerian pressure perturbation: $\ln [ P(\mathbf{r},t) / P_0(\mathbf{r},t)]$
         \\
         $\delta_\rho$
         & Eulerian density perturbation: $\ln [ \rho(\mathbf{r},t) / \rho_0(\mathbf{r},t)]$
         \\
         $\kappa$
         & Spitzer conduction coefficient for diffusion of thermal electrons
         \\
         $\lambda_{\rm F}$
         & Field length: $\sqrt{\kappa T/ n_e \Lambda}$
         \\
         $\lambda_{\rm K}$
         & Entropy scale height: $|\nabla \ln K_0|^{-1}$
         \\
         $\lambda_{\rm P}$
         & Pressure scale height: $|\nabla \ln P_0|^{-1}$
         \\
         $\lambda_{\rm s}$
         & Scale height accounting for vertical gradient in $c_{\rm s}$
         \\
         $\lambda_\Delta$
         & Scale height accounting for Lagrangian compression
         \\
         $\mu$
         & Mean particle mass in units of $m_p$
         \\
         $\boldsymbol{\xi}$
         & Displacement of perturbed flow
         \\
         $\xi_\parallel$
         & Component of displacement $\parallel$ to $\mathbf{\tilde{g}}$
         \\
         $\xi_\perp$
         & Component of displacement $\perp$ to $\mathbf{\tilde{g}}$
         \\
         $\rho$
         & Gas mass density
         \\
         $\rho_0$
         & Unperturbed gas mass density
         \\
         $\omega$
         & Wave frequency
         \\
         $\hat{\omega}$
         & Wave frequency with thermal instability removed
         \\
         $\omega_\pm$
         & Frequencies of thermally unstable gravity-wave modes
         \\
         $\omega_0$
         & Frequency of magnetothermal drip modes
         \\
         $\omega_{\rm A}$
         & Alfv\'en frequency
         \\
         $\omega_{{\rm A},\parallel}$
         & Alfv\'en frequency for waves with displacements $\parallel$ to $\mathbf{\tilde{g}}$
         \\
         $\omega_{{\rm A},\perp}$
         & Alfv\'en frequency for waves with displacements $\perp$ to $\mathbf{\tilde{g}}$
         \\
         $\omega_{\rm ac}$
         & Acoustic cutoff frequency
         \\
         $\hat{\omega}_{\rm ac}$
         & Acoustic cutoff frequency with thermal instability removed
         \\
         $\omega_{\rm buoy}$
         & Frequency of adibatic internal gravity wave
         \\
         $\omega_{\rm BV}$
         & Brunt V\"{a}is\"al\"a frequency for gravitational acceleration $\mathbf{g}$
         \\
         $\tilde{\omega}_{\rm BV}$
         & Brunt V\"{a}is\"al\"a frequency for effective gravitational acceleration $\mathbf{\tilde{g}}$
         \\
         $\omega_{\rm sw}$
         & Growth rate of isentropic pressure perturbations
         \\
         $\omega_{\rm ti}$
         & Growth rate of isobaric entropy perturbations
         \\
        \hline
    \end{tabular}

\newpage

\bibliographystyle{aasjournal}
\bibliography{references}

\end{document}